\documentclass[oneside,american,pdflatex,sn-mathphys-num]{sn-jnl}
\usepackage[T1]{fontenc}
\usepackage[utf8]{inputenc}
\usepackage{colortbl}
\usepackage{booktabs}
\usepackage{bm}
\usepackage{multirow}
\usepackage{amsmath}
\usepackage{amssymb}
\usepackage{graphicx}
\usepackage{esint}

\makeatletter

\providecommand{\tabularnewline}{\\}

\usepackage{multirow}
\providecommand{\tabularnewline}{\\}
\usepackage{babel}
\renewcommand{\bibliographystyle}[1]{}
\newcommand{\vrr}{\vrule width 0pt height 10pt depth 5pt}
\newcommand{\vrt}{\vrule width 0pt height 15pt depth 0pt}
\newcommand{\vrd}{\vrule width 0pt height 0pt depth 15pt}
\newcommand{\vrdd}{\vrule width 0pt height 0pt depth 5pt}
\newcommand{\vrtt}{\vrule width 0pt height 10pt depth 0pt}

\makeatother

\usepackage{babel}
\begin{document}
\title{A generic route to chaos in self-excited plasma oscillations}

\author[1]{\fnm{Bikramjit} \sur{Joardar}}
\author[1]{\fnm{Kalpana} \sur{Bora}}
\author[2]{\fnm{Asheesh} \sur{Gupta}}
\author*[1]{\fnm{Madhurjya P.} \sur{Bora}}\email{mpbora@gauhati.ac.in}

\affil[1]{\orgdiv{Department of Physics}, \orgname{Gauhati University}, 
\orgaddress{\city{Guwahati} \postcode{781014}, \country{India}}}

\affil[2]{\orgdiv{Department of Physics}, \orgname{Galgotias University}, 
\orgaddress{\city{Greater Noida} \postcode{203201}, \country{India}}}

\abstract{We propose a generic autonomous route to chaos in self-excited plasma
oscillations through nonlinear feedback from a finite-rate charging
variable. Near the threshold of an oscillatory ion-acoustic instability,
a multiple-scale reduction yields a Stuart-Landau amplitude equation,
represented by a van der Pol-type modal surrogate and coupled to debris-charge
evolution in a prototype case of plasma-debris interaction scenario.
For a debris-induced ion-ion counter-streaming instability, charge-dependent
forcing and frequency modulation, finite charging relaxation, phase-lagged
modal response, mixed nonlinear feedback, and saturation form a closed
dynamical system without an externally imposed periodic driver. Numerical
bifurcation and Lyapunov analyses reveal multistability and a bounded
chaotic branch reached through period doubling bifurcation and band
formation. We test the proposed mechanism using a one-dimensional
open-boundary \emph{hybrid} particle-in-cell-Monte Carlo collision
simulation incorporating particle collection, photoemission, secondary-electron
emission, and surface regulation. Weak-form sparse regression identifies
finite intervals whose projected dynamics reproduce the sign structure
of the reduced model, while independent $0$-$1$ chaos test, nonlinear
predicatbility, surrogate-data, and Lyapunov diagnostics provide localized
evidence of deterministic chaotic organization in the subsonic case.
The combined results establish a proof-of-principle route by which
a self-excited plasma mode can become chaotic through internally generated
charging feedback.}


\maketitle

\section{Introduction}

The mathematical origin of deterministic chaos can be traced back
to Poincar\'e’s analysis of the three-body problem in 1890 \cite{poincare1892},
although chaos emerged as a distinct and widely studied field only
after Lorenz’s work on a simplified model of the atmosphere in 1963
and the introduction of the term by Li and Yorke in 1975 \cite{lorenz1963,li1975}.
Lorenz’s atmospheric model is an example of so-called “self-excited”
chaos, in the sense that there is no external driving force in the
system and the system, by itself, attains the chaotic state once the
appropriate parameter values are reached. In contrast to this, one
can generally drive a multidimensional system to chaos through an
external driving mechanism. Such externally driven chaos was probably
first observed experimentally by van der Pol and van der Mark in 1927
\cite{vanderpol1927}, first analysed mathematically by Cartwright
and Littlewood in 1945 \cite{cartwright1945}, and then observed in
Ueda’s analog experiments beginning in 1961 and documented in his
subsequent analysis of the Duffing oscillator \cite{ueda1979}.

Self-excited, or autonomous, chaos is quite ubiquitous in nonlinear
dissipative systems, although it requires more structure (dimensions)
than ordinary periodic self-excitation. Apart from the Lorenz system,
the R\"ossler system \cite{rossler1976} and Chua’s circuit model
are some examples of autonomous chaotic systems \cite{matsumoto1985,chua1986}.
While internal coupling of the convection variables leads to chaos
in the Lorenz system, internal nonlinear feedback producing stretching
and folding provides the route to chaos in the R\"ossler system,
whereas in Chua’s circuit it occurs through the nonlinear resistor.
On the other hand, a self-excited oscillator such as the autonomous
van der Pol oscillator,
\begin{equation}
\ddot{x}-\mu(1-x^{2})\dot{x}+\omega^{2}_{0}x=0
\end{equation}
approaches a periodic limit cycle and cannot itself become chaotic
unless a third dynamical variable is introduced that can supply feedback.
Thus, self-excited oscillations are quite common in unstable nonlinear
systems, whereas autonomous chaos generally emerges when the oscillatory
mode is coupled to at least one additional dynamical variable providing
nonlinear feedback or memory. This is exactly what we try to explore
in the present work, where we treat a plasma system as a self-excited
dynamical system and show that it contains the required internal dynamical
elements to attain a chaotic state in the appropriate parameter regime.

\subsection{Difficulty with natural systems}

Observing consistent chaotic dynamics and patterns in natural systems
such as plasmas is quite difficult, as the complete initial state
is never known. A plasma state, for example, requires distribution
functions $f_{s}(\boldsymbol{x},\boldsymbol{v})$, fields, and boundary
conditions -- not merely a few macroscopic variables (the subscript
$s$ here refers to the species). The atmosphere similarly requires
spatially resolved velocity, temperature, pressure, and humidity fields.
In molecular dynamics, the positions and momenta of an enormous number
of particles would be required. Any such unresolved component can
act as an initial-condition uncertainty that is amplified by chaos,
and a consistent chaotic pattern may not emerge even if the underlying
dynamics is chaotic.

Besides, natural systems are open. They continually exchange energy,
particles, and momentum with their surroundings. For example, a natural
plasma experiences changes in the densities and temperatures of the
constituent species, fields, particle injection and loss, radiation,
etc. Consequently, the control parameters themselves change while
the system evolves. The observed signal may therefore come from a
slowly drifting system rather than from a single stationary chaotic
attractor. Apart from this, noise in natural systems and chaos can
look very similar. A broadband spectrum, irregular oscillations, and
an apparently random time series can be produced not only by deterministic
chaos but also by stochastic forcing, instrumental noise, turbulent
superposition, intermittent events, to name a few. Natural systems
are also generally higher-dimensional, making phase-space analysis
and observation extremely difficult. Their dynamics generally involve
multiple timescales, and the overlapping of these timescales makes
long-time, consistent chaotic dynamics extremely difficult to observe.
In short, chaos in natural systems, such as an evolving plasma, is
not necessarily rare, but its unambiguous identification is difficult.

The situation becomes considerably more tractable when the system
is subjected to a controlled external drive. A prescribed driving
amplitude and frequency provide reproducible control parameters and
a natural phase reference, permitting systematic bifurcation scans,
Poincaré sampling, and the identification of recognised routes to
chaos. Consequently, externally driven chaos has been demonstrated
convincingly in a wide variety of nonlinear oscillators and plasma
systems.

\subsection{What are we gong to address?}

A more difficult and fundamental question, which motivates the present
work, is whether a natural system such as a plasma can generate chaos
autonomously, without an imposed time-dependent forcing.

In this work, we address this question by examining whether, near
the threshold of an oscillatory plasma instability, the mode evolution
can generically be represented by a Stuart-Landau amplitude equation
\cite{landau1944,stuart1960,crawford1991}, provided that an isolated
dominant mode and a regular solvability reduction exist. We show that
such a situation exists, especially in a plasma-debris interaction,
which provides us with the required Stuart-Landau amplitude equation
and its van der Pol-type surrogate \cite{vanderpol1926,strogatz2015}.
The latter can acquire autonomous chaotic dynamics through coupling
to a physically evolving internal variable. In the plasma-debris prototype
situation that we consider here, the debris charge supplies this additional
degree of freedom -- instability-induced variations in the local
plasma modify the charging currents, while the evolving charge provides
the feedback mechanism for the mode dynamics. The resulting three-dimensional
autonomous system therefore provides a physically motivated route
to internally generated plasma chaos. The objective is not to claim
that every unstable plasma \emph{must} become chaotic, but to identify
a generic feedback mechanism through which autonomous chaos can arise
and to formulate predictions that can subsequently be tested using
kinetic and particle-in-cell (PIC) simulations \cite{dawson1983,hockney1988,birdsallLangdon1991,verboncoeur2005}.
Our reduced plasma-debris instability model establishes this mathematical
possibility and identifies the necessary feedback terms and dynamical
conditions. PIC simulations are then carried out to determine whether
these terms occur with physically admissible magnitudes and whether
the predicted chaotic basin is realised in the underlying plasma system.It
is important to note that the PIC framework presented here provides
the foundation for a numerical experiment whose physical configuration
can, in principle, be realised through a suitable experimental arrangement.

In Section 2, we describe the Stuart-Landau formalism in the ion-acoustic
regime, where we also reduce the system to a canonical form and develop
the van der Pol surrogate. In Section 3, we adapt our theoretically
developed model to a prototype ion-acoustic instability involving
plasma-debris interaction and the ion-ion counter-streaming instability
(IICSI). In Section 4, we carry out a dynamical analysis of the surrogate
model and show the existence of multistability and a chaotic basin,
assuming a feedback mechanism based on charge fluctuations of the
external debris due to photoemission and secondary electron emission
(SEE) with surface regulation. In Section 5, we describe our open-boundary,
one-dimensional PIC simulation model, which implements these physical
mechanisms within the PIC framework. In Section 6, we briefly describe
how the theory and PIC results are connected, how well the simulation
results reproduce the theoretical predictions, and where the dissimilarities
lie. Finally, in Section 7, we summarise and conclude.

\section{Stuart-Landau equation for a generic ion-acoustic system}

\subsection{The configuration}

We assume our plasma state to be governed by an electrostatic Vlasov-Poisson
system \cite{nicholson1983,krall1973}, with the state variable 
\begin{equation}
F=(f_{i},f_{e}),
\end{equation}
where 
\begin{equation}
\phi={\cal P}[f_{i},f_{e}],
\end{equation}
where ${\cal P}$ denotes the Poisson solver, $f_{i,e}$ are the ion
and electron distribution functions, and $\phi$ is the electrostatic
potential. The governing equations are thus 
\begin{eqnarray}
\frac{\partial f_{s}}{\partial t}+\bm{v}\cdot\nabla f_{s}+\frac{q_{s}}{m_{s}}\bm{E}\cdot\nabla_{v}f_{s} & = & 0,\label{eq:ch11-vlasov}\\
\bm{E} & = & -\nabla\phi,\\
\nabla^{2}\phi & = & -\frac{1}{\epsilon_{0}}\sum_{s=i,e}q_{s}\int f_{s}\,d^{3}v,
\end{eqnarray}
where $s=i,e$ denotes the ions and electrons. We choose an equilibrium
defined by 
\begin{equation}
f_{i}=f_{i0}(v),\quad f_{e}=f_{e0}(v),\quad\phi_{0}=0.
\end{equation}
In the ion-acoustic (IA) regime \cite{fried1961,ichikawa1974}, $f_{e0}$
is usually Maxwellian as electrons are fast with their thermal velocity
which is quite larger than the characteristic IA wave 
\begin{equation}
v_{\textrm{th}e}\gg\frac{\omega}{k}.
\end{equation}
Using a standard perturbation theory to denote the perturbation, we
write 
\begin{equation}
F=F_{0}+\tilde{F},
\end{equation}
with 
\begin{eqnarray}
F_{0} & = & (f_{i0},f_{e0}),\\
\tilde{F} & = & (\tilde{f}_{i},\tilde{f}_{e}).
\end{eqnarray}
The perturbed parts are expressed in a generalised perturbation expansion
\begin{eqnarray}
\tilde{F} & = & \epsilon F_{1}+\epsilon^{2}F_{2}+\epsilon^{3}F_{3}+\cdots,\\
\tilde{\phi} & = & \epsilon\phi_{1}+\epsilon^{2}\phi_{2}+\epsilon^{3}\phi_{3}+\cdots,
\end{eqnarray}
much like the reductive perturbation theory, where $\epsilon$ is
the small expansion parameter. For the purpose of a normal-form reduction,
the perturbed evolution equation can now be written in the generic
form 
\begin{equation}
\partial_{t}\tilde{F}={\cal L}\tilde{F}+{\cal N}_{2}(\tilde{F},\tilde{F})+{\cal N}_{3}(\tilde{F},\tilde{F},\tilde{F})+\cdots,\label{eq:ch11-expansion}
\end{equation}
where ${\cal L}$ is the linear operator and ${\cal N}_{j}$ is the
nonlinear operator at the $j$th order. For the primitive Vlasov--Poisson
system, the nonlinearity is quadratic, so that ${\cal N}_{3}=0$,
though a cubic amplitude term nevertheless appears at third order
through ${\cal N}_{2}(F_{1},F_{2})$. We however retain ${\cal N}_{3}$
in the notation only to allow the discussion to apply to a more general
reduced plasma model. The linear electrostatic kinetic dispersion
relation is given by 
\begin{equation}
D(k,\omega)=0,
\end{equation}
which represents the IA wave with a frequency $\omega_{0}$. We now
consider a weak linear growth rate arising from an intrinsic instability,
so that the complex frequency of the oscillations is given by 
\begin{equation}
\omega=\omega_{0}-i\epsilon^{2}\gamma_{2},\quad\omega_{0}\gg\epsilon^{2}\gamma_{2}>0.
\end{equation}
The subscripts `$0$' and `$2$' are used to denote the order where
these frequencies are considered. Consistently, we separate the operator
into its value at threshold and its small departure from threshold,
\begin{equation}
{\cal L}={\cal L}_{0}+\epsilon^{2}{\cal L}_{2},
\end{equation}
where the leading-order eigenfunction $\varPhi$ satisfies 
\begin{equation}
{\cal L}_{0}\varPhi=i\omega_{0}\varPhi.
\end{equation}
Here $\varPhi$ is the linear eigenfunction or eigenmode structure
at the instability threshold. Away from threshold, the corresponding
modal dependence, with the present sign convention, is 
\begin{equation}
\tilde{F}\sim\varPhi e^{i\omega_{0}t+i\bm{k}\cdot\bm{x}}e^{\epsilon^{2}\gamma_{2}t}+c.c.\label{eq:ch11-modal}
\end{equation}
The above equation can also be physically interpreted as the eigenvalue
equation at the threshold of the instability when $\gamma_{2}=0$.
As the eigenvalue crosses the imaginary axis at this point, it is
called a `Hopf instability point', a terminology borrowed from nonlinear
dynamics \cite{crawford1991,kuznetsov2004}. For a collisionless kinetic
system this reduction additionally assumes that the relevant mode
is sufficiently isolated from the continuous spectrum for a regular
solvability projection to be meaningful.

\subsection{Scale expansion}

We now introduce a slow time $T$ 
\begin{equation}
T=\epsilon^{2}t,
\end{equation}
so that we have 
\begin{equation}
\partial_{t}\to\partial_{t}+\epsilon^{2}\partial_{T}.
\end{equation}

\subsubsection{Order ${\cal O}(\epsilon)$}

At the order ${\cal O}(\epsilon)$, Eq.(\ref{eq:ch11-expansion})
becomes 
\begin{equation}
(\partial_{t}-{\cal L}_{0})F_{1}=0,
\end{equation}
for which we seek a solution of the form 
\begin{equation}
F_{1}\sim\varPhi e^{\lambda t},
\end{equation}
where $\lambda=i\omega_{0}$ is the leading-order eigenvalue. So,
$F_{1}$ has a general form given by 
\begin{equation}
F_{1}=a(T)\varPhi e^{i\omega_{0}t}+a^{*}(T)\varPhi^{*}e^{-i\omega_{0}t},\label{eq:ch11-f1}
\end{equation}
where $a(T)$ is the complex amplitude of the wave. Explicitly 
\begin{eqnarray}
f_{s1} & = & a(T)\varPhi_{s}e^{i\omega_{0}t}+a^{*}(T)\varPhi^{*}_{s}e^{-i\omega_{0}t},\\
\phi_{1} & = & a(T)\varPhi_{\phi}e^{i\omega_{0}t}+a^{*}(T)\varPhi^{*}_{\phi}e^{-i\omega_{0}t}.
\end{eqnarray}
Here $f_{s1}$ is the first-order perturbed distribution function
of species $s$, with $s=i,e$ denoting the ion and electron species,
respectively, and $\phi_{1}$ is the corresponding first-order perturbed
electrostatic potential. Thus $F_{1}=(f_{i1},f_{e1})$, while $\varPhi_{s}$
and $\varPhi_{\phi}$ denote the distribution-function and potential
components of the linear eigenmode, respectively. The potential perturbation
$\phi_{1}$ is determined self-consistently from $f_{i1}$ and $f_{e1}$
through the first-order Poisson equation. The amplitude $a(T)\equiv a$
is a function of the slow timescale and is undetermined at the first
order.

\subsubsection{Order ${\cal O}(\epsilon^{2})$}

At the second order, Eq.(\ref{eq:ch11-expansion}) can be written
as 
\begin{equation}
(\partial_{t}-{\cal L}_{0})F_{2}={\cal N}_{2}(F_{1},F_{1}).\label{eq:ch11-f2}
\end{equation}
Eq.(\ref{eq:ch11-f1}) indicates that the general form for $F_{2}$
is given by (see Appendix) 
\begin{equation}
F_{2}=a^{2}\varPhi_{2}e^{2i\omega_{0}t}+|a|^{2}\varPhi_{0}+(a^{*})^{2}\varPhi_{-2}e^{-2i\omega_{0}t},
\end{equation}
where $\varPhi_{-2}=\varPhi^{*}_{2}$.

We note that, to determine the amplitude $a$, we need some kind of
a solvability condition, which usually comes from a secular term.
We further note that a secular term (or a resonant term) will appear
\emph{only} when we have a harmonic $\propto e^{\pm i\omega_{0}t}$
(see Appendix). As there is no term proportional to $e^{\pm i\omega_{0}t}$
in the expression for $F_{2}$, there cannot be any amplitude equation
at this order and we need to go the third order.

\subsubsection{Order ${\cal O}(\epsilon^{3})$}

At this order, we have the following equation 
\begin{equation}
(\partial_{t}-{\cal L}_{0})F_{3}=-\partial_{T}F_{1}+{\cal L}_{2}F_{1}+2{\cal N}_{2}(F_{1},F_{2})+{\cal N}_{3}(F_{1},F_{1},F_{1}),\label{eq:ch11-f3}
\end{equation}
where for simplicity, we have assumed ${\cal N}_{2}(F_{1},F_{2})$
to be symmetric. From the expressions for $F_{1}$, we find that 
\begin{eqnarray}
-\partial_{T}F_{1} & = & -\dot{a}\varPhi e^{i\omega_{0}t}+c.c.\\
{\cal L}_{2}F_{1} & = & a{\cal L}_{2}\varPhi e^{i\omega_{0}t}+c.c.
\end{eqnarray}
where $\dot{a}=\partial_{T}(a)$. From the expression for $F_{2}$,
we see that in the term ${\cal N}_{2}(F_{1},F_{2})$, the product
$F_{1}F_{2}$ contains terms 
\begin{equation}
\propto a|a|^{2}e^{i\omega_{0}t}+c.c.,\quad a^{3}e^{3i\omega_{0}t}+c.c.
\end{equation}
In the term ${\cal N}_{3}(F_{1},F_{1},F_{1})$, we have 
\begin{equation}
F^{3}_{1}\sim(ae^{i\omega_{0}t}+c.c.)^{3},
\end{equation}
which provides terms 
\begin{equation}
\propto3a|a|^{2}e^{i\omega_{0}t}+c.c.,\quad a^{3}e^{3i\omega_{0}t}+c.c.
\end{equation}
Combining all these, the resonant part at order ${\cal O}(\epsilon^{3})$
can then be written as 
\begin{equation}
R_{3,\textrm{res}}=\left[-\dot{a}\varPhi+a{\cal L}_{2}\varPhi+a|a|^{2}G_{21}\right]e^{i\omega_{0}t}+c.c.,\label{eq:ch11-r3res}
\end{equation}
where $G_{21}$ denote the vector coefficient of $a|a|^{2}e^{i\omega_{0}t}$
generated by the nonlinear terms (see Appendix C),
\begin{equation}
G_{21}=2{\cal N}_{2}(\varPhi,\varPhi_{0})+2{\cal N}_{2}(\varPhi^{*},\varPhi_{2})+3{\cal N}_{3}(\varPhi,\varPhi,\varPhi^{*}).
\end{equation}
For a primitive Vlasov--Poisson system ${\cal N}_{3}=0$ and it reduces
to
\begin{equation}
G_{21}=2{\cal N}_{2}(\varPhi,\varPhi_{0})+2{\cal N}_{2}(\varPhi^{*},\varPhi_{2}).
\end{equation}
 So, the full equation at this order is given by 
\begin{equation}
(\partial_{t}-{\cal L}_{0})F_{3}=R_{3}\equiv R_{3,\textrm{res}}+R_{3,\textrm{nr}},
\end{equation}
where the non-resonant parts are denoted by $R_{3,\textrm{nr}}$,
which contains all other harmonics at the third order. Notice that
the nonlinear resonant vector $G_{21}$ need not be parallel to $\varPhi$,
or in general
\begin{equation}
G_{21}\neq c\varPhi,
\end{equation}
where $c$ is a constant. The consequence is that we cannot factor
out $\varPhi$ from Eq.(\ref{eq:ch11-r3res}) and we must use the
adjoint projection to obtain a scalar coefficient, which is carried
out in the next section.

\subsubsection{The solvability condition and the Stuart-Landau equation}

As the non-resonant terms produce higher-harmonic oscillations without
secular amplitude growth, nothing interesting happens there and we
concentrate on the resonant part of the equation at the third order,
which is given by 
\begin{equation}
{\cal M}F_{3}\equiv(\partial_{t}-{\cal L}_{0})F_{3}=R_{3,\textrm{res}},
\end{equation}
where ${\cal M}\equiv(\partial_{t}-{\cal L}_{0})$ is the operator
which needs to be inverted for a solution 
\begin{equation}
F_{3}={\cal M}^{-1}R_{3,\textrm{res}}.\label{eq:ch11-invert}
\end{equation}
But as we observe from the first order equation that 
\begin{equation}
{\cal M}F_{1}\equiv{\cal M}\left(a\varPhi e^{i\omega_{0}t}\right)=0,
\end{equation}
$\varPhi e^{i\omega_{0}t}$ lies in the nullspace of ${\cal M}$,
which indicates that ${\cal M}$ is singular and Eq.(\ref{eq:ch11-invert})
will not work \cite{kielhofer2012,kuznetsov2004}. In order to seek
a solution, we note that the eigenfunction of ${\cal L}_{0}$ at the
leading order is given by 
\begin{equation}
{\cal L}_{0}\varPhi=i\omega_{0}\varPhi,
\end{equation}
for which we define the adjoint eigenfunction $\Psi$, so that 
\begin{equation}
{\cal L}^{\dag}_{0}\Psi=-i\omega_{0}\Psi,
\end{equation}
with the normalization 
\begin{equation}
\left\langle \Psi,\varPhi\right\rangle =1.
\end{equation}
For the state vectors $A=(A_{i},A_{e})$ and $B=(B_{i},B_{e})$, their
overlap in plasma phase space is defined by
\begin{equation}
\left\langle A,B\right\rangle =\sum_{s=i,e}\int_{\Omega_{x}}d^{3}x\int_{\mathbb{R}^{3}}d^{3}v\,A^{*}_{s}(\bm{x},\bm{v})B_{s}(\bm{x},\bm{v}).
\end{equation}
The sum includes the ion and electron parts of the perturbation. The
integrals are over the complete spatial region and all particle velocities.
For a periodic wave, $\Omega_{x}$ may be chosen as one wavelength
and for a finite system it is the physical plasma domain. The boundary
conditions are assumed not to be affected when an operator is transferred
between the two functions in this integral. The adjoint operator ${\cal L}^{\dag}_{0}$
is simply defined as the operator that satisfies
\begin{equation}
\langle A,{\cal L}_{0}B\rangle=\langle{\cal L}^{\dag}_{0}A,B\rangle.
\end{equation}
Finally, the size of the adjoint eigenfunction is chosen so that $\langle\Psi,\varPhi\rangle=1$.
This normalization ensures that the projection directly extracts the
scalar amplitude of the resonant forcing along the mode $\varPhi$.
We first extract the coefficient of $e^{i\omega_{0}t}$ in Eq.(\ref{eq:ch11-r3res})
and then project that coefficient onto $\Psi$. The solvability condition
is therefore 
\begin{equation}
\left\langle \Psi,-\dot{a}\varPhi+a{\cal L}_{2}\varPhi+a|a|^{2}G_{21}\right\rangle =0.
\end{equation}
Defining 
\begin{equation}
\gamma_{2}=\left\langle \Psi,{\cal L}_{2}\varPhi\right\rangle ,\qquad\nu=\left\langle \Psi,G_{21}\right\rangle ,
\end{equation}
we obtain 
\begin{equation}
-\dot{a}+\gamma_{2}a+\nu|a|^{2}a=0
\end{equation}
for the amplitude $a$. Thus the Landau coefficient $\nu$ is obtained
only after the complete nonlinear resonant forcing has been projected
onto the adjoint mode. Note that in the absence of the third term,
the above equation results in an infinite linear growth of the amplitude
as $\gamma_{2}>0$. So, for nonlinear saturation, we must have 
\begin{equation}
\textrm{Re}(\nu)<0.
\end{equation}
Writing 
\begin{equation}
\nu=-(\ell_{r}+i\ell_{i}),\qquad\ell_{r}>0,
\end{equation}
we have the amplitude equation 
\begin{equation}
\dot{a}=\gamma_{2}a-(\ell_{r}+i\ell_{i})|a|^{2}a.\label{eq:ch11-eqa}
\end{equation}
Here $\ell_{r}$ produces nonlinear saturation, whereas $\ell_{i}$
produces a nonlinear frequency shift. Note that this equation is \emph{only}
for the amplitude evolution in the slow time scale $T$. For the complete
oscillation scenario, we need to include IA oscillations with its
frequency $\omega_{0}$ and thus define the amplitude of the complete
IA oscillation as $A(t)$, 
\begin{equation}
A(t)=a(T)e^{i\omega_{0}t},
\end{equation}
so that 
\begin{equation}
\frac{dA}{dt}=i\omega_{0}A+\epsilon^{2}\dot{a}e^{i\omega_{0}t},
\end{equation}
where $da/dt=\epsilon^{2}\dot{a}$. Substituting for $\dot{a}$ from
Eq.(\ref{eq:ch11-eqa}), we have 
\begin{equation}
\frac{dA}{dt}=i\omega_{0}A+\epsilon^{2}\left[\gamma_{2}A-(\ell_{r}+i\ell_{i})|A|^{2}A\right],
\end{equation}
which is the Stuart-Landau (SL) equation for the IA mode \cite{landau1944,stuart1960,crawford1991}.
Absorbing the factor $\epsilon^{2}$ into the slow coefficients and
writing the physical growth rate as $\gamma$, the SL equation becomes
\begin{equation}
\frac{dA}{dt}=(\gamma+i\omega_{0})A-(\ell_{r}+i\ell_{i})|A|^{2}A,\label{eq:ch11-sl}
\end{equation}
with $A(t)$ as the complex modal amplitude.

\subsection{Reduction to a canonical form}

We now decompose the complex modal amplitude $A(t)$ into a physical
amplitude and the phase part 
\begin{equation}
A=\frac{1}{2}R(t)e^{i\theta(t)},
\end{equation}
with $|A|^{2}=R^{2}/4$. Hence 
\begin{equation}
\frac{dA}{dt}=\frac{1}{2}\left(\frac{dR}{dt}+iR\frac{d\theta}{dt}\right)e^{i\theta}.
\end{equation}
Substituting these in Eq.(\ref{eq:ch11-sl}), we have 
\begin{equation}
\frac{dR}{dt}+iR\frac{d\theta}{dt}=\gamma R-\ell_{r}\frac{R^{3}}{4}+i\left(\omega_{0}R-\ell_{i}\frac{R^{3}}{4}\right).\label{eq:ch11-sl-1}
\end{equation}
Comparing the real and imaginary parts, 
\begin{eqnarray}
\frac{dR}{dt} & = & \gamma R-\ell_{r}\frac{R^{3}}{4},\label{eq:ch11-ampl1}\\
\frac{d\theta}{dt} & = & \omega_{0}-\ell_{i}\frac{R^{2}}{4}.\label{eq:ch11-theta}
\end{eqnarray}
From the amplitude equation (\ref{eq:ch11-ampl1}), we see that 
\begin{equation}
\begin{array}{rcll}
{\displaystyle \frac{dR}{dt}} & \simeq & \gamma R, & \quad\textrm{for small}\,R,\\
 & \sim & -{\displaystyle \frac{\ell_{r}}{4}R^{3},} & \quad\textrm{for large}\,R,
\end{array}
\end{equation}
which shows that small-amplitude oscillations grow while sufficiently
large-amplitude oscillations are suppressed, a typical behaviour of
limit-cycle oscillators like the van der Pol-type oscillators \cite{vanderpol1926,strogatz2015}.
The radial amplitude dynamics in Eq.(\ref{eq:ch11-ampl1}) has the
characteristic structure of a self-excited oscillator -- small-amplitude
perturbations grow, whereas sufficiently large amplitudes are nonlinearly
suppressed. We therefore seek a real second-order oscillator whose
weakly nonlinear averaged amplitude dynamics reproduces Eq.(\ref{eq:ch11-ampl1}).
A convenient choice is the van der Pol-type surrogate
\begin{equation}
\ddot{x}-\mu(1-\alpha x^{2})\dot{x}+\omega^{2}_{0}x=0.\label{eq:ch11-vd}
\end{equation}
We should note that the above form is by no means unique. It is a
convenient real second-order surrogate whose averaged radial-amplitude
dynamics reproduces the real, saturating part of the SL amplitude
equation (see below). It does not reproduce the nonlinear frequency
shift proportional to $\ell_{i}$ unless an additional reactive term
is introduced.

For weak nonlinearity, assuming $\mu\ll\omega_{0}$, the above equation
can be approximated as 
\begin{equation}
\ddot{x}=-\omega^{2}_{0}x.
\end{equation}
So, to the lowest order the motion is nearly harmonic and we also
assume that the amplitude $x$ with a nearly harmonic motion 
\begin{equation}
x=R(t)\,\cos\psi,\quad\psi=\omega_{0}t+\vartheta(t),\label{eq:ch11-x}
\end{equation}
with $R$ and $\vartheta$ varying slowly in time. Up to the leading
order, 
\begin{equation}
\dot{x}=-\omega_{0}R\,\sin\psi,\label{eq:ch11-xdot}
\end{equation}
with $\dot{x}\equiv dx/dt$. We now define the oscillator energy ${\cal E}$
as 
\begin{equation}
{\cal E}=\frac{1}{2}\dot{x}^{2}+\frac{1}{2}\omega^{2}_{0}x^{2}
\end{equation}
and its time derivative 
\begin{equation}
\dot{{\cal E}}=\dot{x}\left(\ddot{x}+\omega^{2}_{0}x\right).
\end{equation}
Writing the expression $\left(\ddot{x}+\omega^{2}_{0}x\right)$ in
terms of $\dot{{\cal E}}$ in the oscillator equation (\ref{eq:ch11-vd}),
we get 
\begin{equation}
\dot{{\cal E}}=\mu(1-\alpha x^{2})\dot{x}^{2}.
\end{equation}
Substituting for $x$ and $\dot{x}$ from Eqs.(\ref{eq:ch11-x},\ref{eq:ch11-xdot}),
we get 
\begin{equation}
\dot{{\cal E}}=\mu(1-\alpha R^{2}\cos^{2}\psi)\omega^{2}_{0}R^{2}\sin^{2}\psi.
\end{equation}
Averaging over one period, we get 
\begin{equation}
\left\langle \dot{{\cal E}}\right\rangle =\mu\omega^{2}_{0}R^{2}\left(\frac{1}{2}-\alpha\frac{R^{2}}{8}\right).\label{eq:ch11-dote}
\end{equation}
For a nearly harmonic motion we have 
\begin{equation}
{\cal E}\simeq\frac{1}{2}\omega^{2}_{0}R^{2}\,\sin^{2}\psi+\frac{1}{2}\omega^{2}_{0}R^{2}\cos^{2}\psi=\frac{1}{2}\omega^{2}_{0}R^{2}
\end{equation}
so that we have 
\begin{equation}
\left\langle \dot{{\cal E}}\right\rangle \equiv\dot{{\cal E}}\simeq\omega^{2}_{0}R\dot{R}.
\end{equation}
Using Eq.(\ref{eq:ch11-dote}), we have 
\begin{equation}
\dot{R}=\frac{\mu}{2}R-\frac{\alpha\mu}{8}R^{3}.
\end{equation}
Comparing the above amplitude equation and Eq.(\ref{eq:ch11-ampl1}),
substituting 
\begin{equation}
\mu=2\gamma,\quad\alpha=\frac{\ell_{r}}{\gamma},
\end{equation}
Eq.(\ref{eq:ch11-vd}) can be written as 
\begin{equation}
\ddot{x}-2\gamma\left(1-\frac{\ell_{r}}{\gamma}x^{2}\right)\dot{x}+\omega^{2}_{0}x=0.\label{eq:ch11-vd1}
\end{equation}
Introducing a re-scaled amplitude 
\begin{equation}
\xi=x\sqrt{\ell_{r}/\gamma},
\end{equation}
Eq.(\ref{eq:ch11-vd1}) can be written in the canonical van der Pol
form 
\begin{equation}
\ddot{\xi}-\mu(1-\xi^{2})\dot{\xi}+\omega^{2}_{0}\xi=0.\label{eq:ch11-vanderpol}
\end{equation}
Thus, provided that an isolated IA mode admits a regular weakly nonlinear
reduction near threshold, its modal amplitude can be described by
a generic SL equation. The van der Pol equation is not an exact reduction
of the kinetic system; rather, it is a convenient real surrogate with
the same leading-order averaged radial-amplitude dynamics. In the
next section, we adapt this surrogate to a plasma-debris configuration
with addition of a feedback mechanism providing us dynamical equation
capable of leading us to chaos.

\section{A prototype -- plasma-debris interaction}

We now discuss the above formalism in the light of the plasma-debris
interaction. It has been already shown that a negatively charged external
debris excites the so-called \emph{self-excited} ion-ion counter-streaming
instability (IICSI), which may result in the formation of pinned solitons
in the nonlinearly saturated regime of the instability \cite{sen2015,arora2021,das2025}.
Near the instability threshold, and provided that a single oscillatory
mode dominates, this behavior motivates the SL description developed
above. In the van der Pol surrogate, $\mu=2\gamma$ is therefore associated
with the linear IICSI growth rate, whereas the nonlinear damping models
its saturation. This identification is a reduced-model hypothesis
and the coefficients must ultimately be determined from kinetic theory
or PIC data.

\subsection{Self-excited IICSI}

The linear theory for IICSI during an electrostatic plasma-debris
interaction can be built up with the 1D Vlasov equation for the ions
\cite{das2025,nicholson1983}
\begin{equation}
\frac{\partial f_{i}}{\partial t}+v\frac{\partial f_{i}}{\partial x}+\frac{e}{m_{i}}E\frac{\partial f_{i}}{\partial v}=0,
\end{equation}
where $f_{i}$ is ion velocity distribution, $m_{i}$ is the ion mass,
and $E=-\partial\phi/\partial x$ is the electric field with $\phi$
being the plasma potential. For the specific IICSI prototype, we adopt
the reduced kinetic-ion and Boltzmann-electron description. With the
perturbation $\sim e^{i(kx-\omega t)}$, the kinetic dispersion relation
can be written as 
\begin{equation}
\varepsilon(k,\omega)=1+\frac{1}{k^{2}\lambda^{2}_{De}}+\frac{\omega^{2}_{\textrm{pi}}}{k^{2}n_{0}}\int_{C}\frac{\partial f_{i0}/\partial v}{\omega/k-v}\,dv=0,
\end{equation}
where 
\begin{equation}
\lambda_{De}=\left(\frac{\epsilon_{0}T_{e}}{n_{0}e^{2}}\right)^{1/2},\quad\omega_{\textrm{pi}}=\left(\frac{n_{0}e^{2}}{m_{i}\epsilon_{0}}\right)^{1/2}.
\end{equation}
Here $\lambda_{De}$ is the electron Debye length, $\omega_{\textrm{pi}}$
is the ion plasma frequency and $C$ denotes the Landau contour. The
subscript `$0$' indicates equilibrium values of the respective quantities.
In deriving the above dispersion relation, the temperature is expressed
in energy units.

As the debris potential (created at its boundary) is screened within
a very thin layer $\delta\sim\lambda_{De}\ll a$, each side of the
debris develops a predominantly local sheath. The overlap of ions
accelerated from opposite sides is therefore reduced. Efficient overlap
is expected when $a\lesssim\lambda_{De}$, so that the sheath extends
over a distance comparable to the debris size. It can be shown that
the necessary conditions for developing a self-excited IICSI for a
spherical debris are \cite{das2025}
\begin{equation}
\begin{array}{rcl}
\phi_{d} & < & 0,\\
|\phi_{d}| & > & |\phi_{d}|_{\textrm{crit}},\\
a & \lesssim & \lambda_{De},
\end{array}
\end{equation}
where the critical debris potential for excitation of the IICSI is
given by
\begin{equation}
|\phi_{d}|_{\textrm{crit}}=\frac{\eta^{2}_{\mathrm{IICSI}}}{8e}T_{i}
\end{equation}
and $\eta_{\mathrm{IICSI}}$ is the dimensionless velocity-separation
factor entering the IICSI threshold estimate, typically of order $5-10$.
The subscript distinguishes it from the regression constraint margin
introduced later.

\subsection{Charging feedback}

Let us assume that the debris charge can fluctuate because of its
interaction with the flowing plasma or because of external effects
such as photoemission following photo-ionising radiation and also
due to secondary electron emissions (SEE) \cite{whipple1981,goertz1989,shukla2002}.
So, in general, we can model the time dependence of debris charge
as 
\begin{equation}
Q_{d}(t)=Q_{d0}+q(t),
\end{equation}
where $Q_{d}(t)$ is the time-dependent debris charge and is assumed
to consist of a base charge $Q_{d0}$ and a fluctuating part $q(t)$,
so that the total current $I_{\textrm{tot}}$ to the debris without
the fluctuating part is zero
\begin{equation}
I_{\textrm{tot}}(Q_{d0})=0.
\end{equation}
 Writing the van der Pol surrogate Eq.(\ref{eq:ch11-vanderpol}) without
any feedback term as
\begin{equation}
\ddot{\xi}=\mu(1-\xi^{2})\dot{\xi}-\omega^{2}_{0}\xi,
\end{equation}
we realize that $-\omega^{2}_{0}\xi$ is the linear restoring force
acting on the mode coordinate $\xi$. So, with the nonlinear feedback
term, we can write the above equation as
\begin{equation}
\ddot{\xi}=\mu\left(1-\xi^{2}\right)\dot{\xi}+F(\xi).\label{eq:vdp_force_form}
\end{equation}
For only linear forcing, $F(\xi)\equiv F_{0}(\xi)=-\omega^{2}_{0}\xi$.
Subsequently this linear part can also be represented locally by an
effective potential $U_{0}(\xi)$,
\begin{equation}
F_{0}(\xi)=-\frac{\mathrm{d}U_{0}}{\mathrm{d}\xi},\qquad U_{0}(\xi)=\frac{1}{2}\omega^{2}_{0}\xi^{2},\label{eq:unperturbed_force_potential}
\end{equation}
A nonzero $q(t)$ modifies the self-consistent electrostatic field
and therefore the effective restoring force acting on the ion-acoustic
mode. We represent the locally conservative part of this interaction
by an effective potential $U(\xi,q)$ satisfying 
\begin{equation}
F(\xi,q)=-\frac{\partial U(\xi,q)}{\partial\xi},\qquad U(\xi,0)=U_{0}(\xi).\label{eq:charge_dependent_potential}
\end{equation}
The use of an effective potential here applies only to the conservative
part of the reduced mode dynamics. Assuming the electrostatic response
to be smooth in the vicinity of $q(t)=0$, the potential may be expanded
to first order in the charge perturbation, 
\begin{equation}
U(\xi,q)=U_{0}(\xi)+qU_{1}(\xi)+\mathcal{O}(q^{2}),\label{eq:potential_charge_expansion}
\end{equation}
where 
\begin{equation}
U_{1}(\xi)=\left.\frac{\partial U(\xi,q)}{\partial q}\right|_{q=0}\label{eq:U1_definition}
\end{equation}
is the leading variation of the mode potential per unit charge perturbation.
A further local expansion of $U_{1}$ about the reference mode state
$\xi=0$ gives 
\begin{equation}
U_{1}(\xi)=U_{1}(0)+U_{1}'(0)\xi+\frac{1}{2}U_{1}''(0)\xi^{2}+\mathcal{O}(\xi^{3}).\label{eq:U1_expansion}
\end{equation}
The term $qU_{1}(0)$ is independent of $\xi$ and therefore makes
no contribution to the generalized force. We consequently define the
two leading charge-response coefficients as 
\begin{equation}
\kappa=-U_{1}'(0),\qquad\beta=U_{1}''(0).\label{eq:kappa_beta_definition}
\end{equation}
With these definitions, the leading charge-induced correction becomes
\begin{equation}
U_{1}(\xi)=-\kappa\xi+\frac{1}{2}\beta\xi^{2}+\mathcal{O}(\xi^{3}),\label{eq:U1_minimal}
\end{equation}
and the effective potential becomes
\begin{equation}
U(\xi,q)=\frac{1}{2}\omega^{2}_{0}\xi^{2}-\kappa q\xi+\frac{1}{2}\beta q\xi^{2}+\mathcal{O}\left(q^{2},q\xi^{3}\right)\label{eq:effective_potential}
\end{equation}
and the corresponding generalized force is given by
\begin{align}
F(\xi,q) & =-\frac{\partial U}{\partial\xi}=-\left(\omega^{2}_{0}+\beta q\right)\xi+\kappa q+\mathcal{O}\left(q^{2},q\xi^{2}\right)
\end{align}
Substituting into Eq.(\ref{eq:vdp_force_form}) and neglecting the
higher-order terms, yields 
\begin{equation}
\ddot{\xi}-\mu\left(1-\xi^{2}\right)\dot{\xi}+\left(\omega^{2}_{0}+\beta q\right)\xi=\kappa q,\qquad q\equiv q(t),\label{eq:vdp_charge_feedback}
\end{equation}
where the two coupling coefficients describe physically distinct charge-feedback
channels. The coefficient 
\begin{equation}
\kappa=-\left.\frac{\partial^{2}U}{\partial\xi\,\partial q}\right|_{\xi=q=0}=\left.\frac{\partial F}{\partial q}\right|_{\xi=q=0}\label{eq:kappa_response}
\end{equation}
measures the displacement-independent change of the generalized force
term caused by a charge perturbation. It therefore represents a charge-induced
shift of the mode's force balance. Such a contribution is possible
when reflection symmetry is broken, as occurs naturally in a flowing
plasma, containing a localized charged object and an asymmetric downstream
wake {[}?{]}. On the other hand, 
\begin{equation}
\beta=\left.\frac{\partial^{3}U}{\partial\xi^{2}\partial q}\right|_{\xi=q=0}=-\left.\frac{\partial^{2}F}{\partial\xi\,\partial q}\right|_{\xi=q=0}\label{eq:beta_response}
\end{equation}
measures the charge dependence of the linear restoring coefficient.
Eq.\eqref{eq:vdp_charge_feedback} should therefore be understood
as a leading-order response model rather than as a consequence of
the isolated van der Pol equation alone. Its construction assumes
that $q$ is a sufficiently small departure from the reference charge,
so that terms of order $q^{2}$ may be neglected, and that the charge-linear
correction to the effective potential can be truncated at quadratic
order in $\xi$. 

As in our intended PIC experiment, $q(t)$ can be directly observed
along with $\xi(t)$, we can straightforwardly do a data-driven test
to see if Eq.(\ref{eq:vdp_charge_feedback}) is satisfied. However,
this procedure provides only a charge-driven model for $\xi$, it
does not describe how $q$ evolves in a new trajectory. To obtain
a model capable of predicting future trajectories, bifurcations and
chaos, one needs a closure for $\dot{q}$.

\subsection{A non-autonomous closure}

We note that the mode evolution equation with charging feedback as
given by Eq.(\ref{eq:vdp_charge_feedback}) needs a closure equation
for $q(t)$. The easiest possible route is a phenomenological approximation,
assuming that the charge signal contains one dominant nearly periodic
component, possibly imposed by an external driver 
\begin{equation}
q(t)=\hat{q}\,\cos(\Omega t+\delta).
\end{equation}
This is the kind of scenario discussed in a model by Shukla \cite{momeni2007}.
Though one can raise very many questions about validity of such an
assumption, we nevertheless consider this case as a possibility. Substituting
it in Eq.(\ref{eq:vdp_charge_feedback}), we get an equation which
has the form of a forced van der Pol-Mathieu (f-vdPM) equation \cite{momeni2007,boraSarmah2007}
\begin{equation}
\ddot{\xi}-\mu(1-\xi^{2})\dot{\xi}+\omega^{2}_{0}\left[1+h\cos(\Omega t+\delta)\right]\xi=g\,\cos(\Omega t+\delta),
\end{equation}
where $h=\beta\hat{q}/\omega^{2}_{0}$ and $g=\kappa\hat{q}$.

The periodically modulated vdPM equation is known to possess periodic,
quasiperiodic and chaotic regimes in different parts of its parameter
space. The above equation therefore identifies a \emph{possible route}
to chaos through charge-induced parametric modulation. It does not,
however, prove that the plasma-debris system is chaotic without the
imposition of the external driving frequency $\Omega$, which is also
its biggest drawback. There are works following this route in various
plasma configurations where an external driving source is introduced
and these systems are proven to lead to chaotic regimes. The purpose
of the present work is, however, to see if the plasma dynamics can
inherently contain enough ingredients to manifest a chaotic component,
however small that may be. Toward this, we consider the self-excited
IICSI that can naturally arise during plasma-debris interactions.

\subsection{An autonomous closure}

The charging feedback analysis outlined in the previous subsection
shows that the instability itself can modify the local plasma density,
potential and the electron and ion velocity distributions. These variations
in turn alter electron and ion currents to the debris. In addition
to this, photoemission or SEE-induced currents can play a vital role
in the evolution of the entire dynamics, when they are physically
relevant. The resulting charge variation changes the debris sheath,
ion acceleration and the counter-streaming configuration, creating
a feedback loop. This helps us in writing a closure equation for $q(t)$
as
\begin{equation}
\frac{dq}{dt}=I_{e}+I_{i}+I_{\mathrm{ph}}+I_{\mathrm{SEE}}+I_{\mathrm{rec}}\cdots=I_{\textrm{tot}},
\end{equation}
These $I_{\alpha}\equiv I_{e,i,\mathrm{ph},\mathrm{SEE},\dots}$ are
current to debris and are functions of debris charge and local plasma
variables 
\begin{equation}
I_{\alpha}=I_{\alpha}(Q_{d},n_{e},n_{i},\phi,u_{i},T_{e},T_{i},Y_{\mathrm{ph}},\varPhi_{\gamma},\dots),
\end{equation}
where $Y_{\mathrm{ph}},\varPhi_{\gamma}$ are photoelectric yield
and photon flux. In the above expression $I_{\mathrm{ph},\mathrm{SEE},\mathrm{rec}}$
are the photoemission, SEE, and recollection currents to the debris.
Assuming that the plasma with self-excited IICSI has reached a statistically
stationary state, we define $Q_{d0}$ and the corresponding plasma
quantities $X_{0}$ by their mean values in this state. The mean charging-current
balance then satisfies $I_{\textrm{tot}}|_{0}=\sum_{\alpha}I_{\alpha}|_{0}=0$.
Expanding the instantaneous total current about this statistical reference
state, we obtain
\begin{equation}
\dot{q}=qI_{Q}+\sum_{X}I_{X}\,\delta X+\frac{1}{2}q^{2}I_{QQ}+q\sum_{X}I_{QX}\,\delta X+\frac{1}{6}q^{3}I_{QQQ}+{\cal R},\label{eq:qexpansion}
\end{equation}
where ${\cal R}$ is rest of the expansion and $I_{Q,X,QX,QQQ}$ denotes
the derivatives of the total current at the reference state 
\begin{equation}
I_{Q}=\left.\frac{\partial I_{\textrm{tot}}}{\partial Q_{d}}\right|_{0},\qquad I_{X}=\left.\frac{\partial I_{\textrm{tot}}}{\partial X}\right|_{0}.
\end{equation}
The quantity $X$ represents various plasma variables
\begin{equation}
X\in\{n_{e},n_{i},\phi,u_{i},T_{e},T_{i},\dots\},
\end{equation}
so that
\begin{equation}
\sum_{X}I_{X}\,\delta X=\left.\frac{\partial I_{\textrm{tot}}}{\partial n_{e}}\right|_{0}\delta n_{e}+\left.\frac{\partial I_{\textrm{tot}}}{\partial n_{i}}\right|_{0}\delta n_{i}+\left.\frac{\partial I_{\textrm{tot}}}{\partial\phi}\right|_{0}\delta\phi+\left.\frac{\partial I_{\textrm{tot}}}{\partial u_{i}}\right|_{0}\delta u_{i}+\cdots.
\end{equation}
Writing
\begin{equation}
\nu_{\textrm{ch}}=-\left.\frac{\partial I_{\textrm{tot}}}{\partial Q_{d}}\right|_{0},\quad h=\frac{1}{2}I_{QQ},\quad g=-\frac{1}{6}I_{QQQ}.
\end{equation}
and neglecting ${\cal R}$, the complete dynamical model can be written
as
\begin{eqnarray}
\ddot{\xi}-\mu\left(1-\xi^{2}\right)\dot{\xi}+\left(\omega^{2}_{0}+\beta q\right)\xi & = & \kappa q,\label{eq:evol1}\\
\dot{q}+\nu_{\textrm{ch}}q & = & \sum_{X}I_{X}\,\delta X+hq^{2}+q\sum_{X}I_{QX}\,\delta X-gq^{3},\label{eq:evol2}
\end{eqnarray}
where $\nu_{\textrm{ch}}$ is the charge relaxation rate of the debris.
The minus sign in the definition of $g$ follows from writing the
cubic Taylor term as $-gq^{3}$. Thus, $g>0$ corresponds to the locally
stabilizing case $I_{QQQ}<0$. For a stable charging equilibrium,
$\nu_{\textrm{ch}}>0$, so that in the absence of perturbations a
small charge disturbance $q(t)$ relaxes exponentially toward $q(t)=0$,
with charging relaxation time $\tau_{\textrm{ch}}=1/\nu_{\textrm{ch}}$.

\subsubsection{Projection onto the IICSI mode}

Assuming the dominant IICSI mode to be represented by the two-dimensional
modal state $(\xi,\dot{\xi})$, the component of each plasma observable
associated with this mode can be projected onto this pair. So, we
can map the right hand side of Eq.(\ref{eq:evol2}) in terms of the
mode of the instability in terms of $(\xi,\dot{\xi})$. To the leading
order, we can write
\begin{equation}
\delta X(t)=A_{X}\xi(t)+B_{X}\dot{\xi}(t)+\delta X_{\perp}(t),\label{eq:expansion}
\end{equation}
where $(A_{X},B_{X})$ are coefficients of $\xi$ and $\dot{\xi}$,
and $\delta X_{\perp}(t)$ contains the remainder of the expansion.
See Appendix D for explicit expressions for $(A_{X},B_{X})$ and a
physical interpretation of $\delta X_{\perp}(t)$. For nearly harmonic
motion, we can write
\begin{equation}
\xi(t)=R\,\cos(\omega t+\theta),\qquad\dot{\xi}(t)=-\omega R\,\sin(\omega t+\theta).
\end{equation}
The pair $(\xi,\dot{\xi})$ therefore spans the two quadratures of
the dominant oscillation. So, any observable oscillating at the same
frequency with an arbitrary phase can be represented by a suitable
linear combination of these two signals. Substituting Eq.(\ref{eq:expansion})
into Eq.(\ref{eq:evol2}), and defining
\begin{equation}
\begin{array}{rclcrcl}
a & = & \sum_{X}I_{X}A_{X}, & \quad & b & = & \sum_{X}I_{X}B_{X},\\
\\c & = & \sum_{X}I_{QX}A_{X}, & \quad & d & = & \sum_{X}I_{QX}B_{X},
\end{array}
\end{equation}
and neglecting the residual contribution, we can write Eq.(\ref{eq:evol2})
as
\begin{equation}
\dot{q}+\nu_{\textrm{ch}}q=a\xi+b\dot{\xi}+hq^{2}+c\xi q+d\dot{\xi}q-gq^{3}.\label{eq:evol21}
\end{equation}

\subsubsection{Minimal normal form}

We now consider the reduced minimal model which has the smallest autonomous
system containing primarily four physical effects -- (\emph{i}) linear
charging relaxation, (\emph{ii}) forcing of charge by the IICSI phase-space
state, (\emph{iii}) modulation of the charging rate by the instantaneous
IICSI state, and (\emph{iv}) bounded nonlinear saturation of large
charge perturbations. These physical effects are supplied by the following
terms, respectively,
\begin{equation}
-\nu_{\mathrm{ch}}q,\qquad a\xi+b\dot{\xi},\qquad c\xi q,\qquad-gq^{3}\quad(g>0).
\end{equation}
Assuming the in-phase mixed response to be dominating the quadrature
mixed response, we can assume $|d\dot{\xi}q|\ll|c\xi q|$, and the
 minimal charging closure becomes
\begin{equation}
\dot{q}+\nu_{\mathrm{ch}}q=a\xi+b\dot{\xi}+hq^{2}+c\xi q-gq^{3}.\label{eq:minimal-charge}
\end{equation}
Among all the terms on the right hand side of the above equation,
we note that $q^{2}$ has the same sign for $q_{0}\gtrless0$ and
so $hq^{2}$ introduces a directional asymmetry and disturbs the nonlinear
current balance. As the total current satisfies
\begin{equation}
I_{\mathrm{tot}}(Q_{d0})=0,
\end{equation}
at the reference state, near this state we have,
\begin{equation}
I_{\mathrm{tot}}(Q_{d0}+q)=-\nu_{\mathrm{ch}}q+hq^{2}-gq^{3}+\cdots,
\end{equation}
where we recognize the quadratic coefficient is $h$ to be
\begin{equation}
h=\frac{1}{2}I_{\mathrm{tot}}''(Q_{d0})\equiv\frac{1}{2}I_{QQ}.
\end{equation}
We note that the reference state represented by $Q_{d0}$ is established
by the competition between currents with different signs. For example,
an increasingly negative charge suppresses electron collection, while
modifying ion attraction in the opposite manner. This inhibits escape
and enhances recollection of emitted electrons. Similarly, SEE and
photoemission contribute positive charging currents which also change
with surface potential \cite{draine1979,chow1993,kimura2014}. Though
these effects need not be individually symmetric, the net current
may be weak
\begin{equation}
I_{e,QQ}+I_{i,QQ}+I_{\mathrm{ph},QQ}+I_{\mathrm{SEE},QQ}+I_{\mathrm{rec},QQ}\simeq0,\label{eq:assump}
\end{equation}
where $I_{\mathrm{rec}}$ is the recollection current. Therefore,
\begin{equation}
I_{\mathrm{tot}}''(Q_{d0})\simeq0\Rightarrow h\simeq0,
\end{equation}
so that the minimal normal dynamical form of the charge evolution
equation becomes
\begin{equation}
\dot{q}+\nu_{\textrm{ch}}q=a\xi+b\dot{\xi}+c\xi q-gq^{3}.\label{eq:evol3}
\end{equation}
Together with Eq.(\ref{eq:evol1}), the above equation now describes
the complete evolution equations used for dynamical analysis and for
comparison with the PIC experiment.

We note that Eq.(\ref{eq:assump}) leading to the final form of the
charge evolution equation is not rigorously derived, it is based on
\emph{a priori} plausible assumption, which is however supported by
the PIC simulation, as we shall see.

\section{Dynamical analysis of the surrogate model}

We now analyze the Stuart-Landau-based van der Pol surrogate model,
namely Eqs.(\ref{eq:evol1},\ref{eq:evol3})
\begin{eqnarray}
\ddot{\xi}-\mu\left(1-\xi^{2}\right)\dot{\xi}+\left(\omega^{2}_{0}+\beta q\right)\xi & = & \kappa q,\label{eq:evol1-1}\\
\dot{q}+\nu_{\textrm{ch}}q & = & a\xi+b\dot{\xi}+c\xi q-gq^{3}.\label{eq:evol2-1}
\end{eqnarray}
Introducing the following dimensionless variables
\begin{equation}
\begin{array}{rclcrcl}
\tau & = & \omega_{0}t, & \qquad & z & = & q/Q_{0},\\
\epsilon & = & \mu/\omega_{0}, &  & \Lambda & = & \nu_{\textrm{ch}}/\omega_{0},\\
H & = & \beta Q_{0}/\omega^{2}_{0}, &  & K & = & \kappa Q_{0}/\omega^{2}_{0},\\
C & = & a/(\omega_{0}Q_{0}), &  & D & = & b/Q_{0},\\
E & = & c/\omega_{0}, &  & G & = & gQ^{2}_{0}/\omega_{0},
\end{array}
\end{equation}
where $Q_{0}>0$ is a characteristic charge scale used to normalize
the charge perturbation and is not the equilibrium debris charge $Q_{d0}$.
The normalized autonomous dynamical model can now be written as 
\begin{align}
\xi' & =y,\label{eq:xi}\\
y' & =\epsilon(1-\xi^{2})y-(1+Hz)\xi+Kz,\label{eq:y}\\
z' & =C\xi+Dy-\Lambda z+E\xi z-Gz^{3},\label{eq:z}
\end{align}
where $\xi$ is a narrow-band collective plasma observable, $y=\xi'$
is its dimensionless velocity or quadrature, and $z$ is the normalized
charge perturbation.
\begin{table}
\caption{Parameter values for chaotic basin.\vrdd}\label{tab:Parameter-values-for}

\begin{tabular}{ccccccc}
\hline 
Parameter & \quad{} & Range & \quad{} & Retained value & \quad{} & \raggedright{}Interpretation\vrr\tabularnewline
\hline 
\hline 
$\epsilon$ &  & $0.02$ to $0.3$ &  & $0.2611$ &  & \vrt%
\parbox[c]{4cm}{%
\raggedright{} Weak to moderately weak oscillator growth%
}\vrd\tabularnewline
$\Lambda$ &  & $0.1$ to $10$ &  & $0.1907$ &  & %
\parbox[c]{4cm}{%
\raggedright{} Use logarithmic spacing%
}\vrd\tabularnewline
$H$ &  & $-0.5$ to $0.5$ &  & $0.03416$ &  & %
\parbox[c]{4cm}{%
\raggedright{} Restoring-force feedback%
}\vrd\tabularnewline
$K$ &  & $-0.5$ to $0.5$ &  & $-0.3156$ &  & %
\parbox[c]{4cm}{%
\raggedright{} Direct electric feedback%
}\vrd\tabularnewline
$C$ and $D$ &  & $-3$ to $3$ each &  & $-2.3761$ and $-0.9267$ &  & %
\parbox[c]{4cm}{%
\raggedright{}In-phase and quadrature charging-current projections%
}\vrd\tabularnewline
$E$ &  & $-1.0$ to $0$ &  & $-0.7871$\vrdd &  & \multirow{2}{*}{Charge relaxation}\tabularnewline
$G$ &  & $0$ to $1.5$ &  & $1.1833$ &  & \tabularnewline
\hline 
\end{tabular}
\end{table}

\subsection{Multistability and chaotic basin}

The theoretical model represented by Eqs.(\ref{eq:xi}-\ref{eq:z})
is integrated by a fourth-order Runge-Kutta solver. For state $\bm{u}=(\xi,y,z)^{T}$,
the Jacobian 
\begin{equation}
J=\begin{pmatrix}0 & 1 & 0\\
-1-Hz-2\epsilon\xi y & \epsilon(1-\xi^{2}) & K-H\xi\\
C+Ez & D & -\Lambda+E\xi-3Gz^{2}
\end{pmatrix}
\end{equation}
is used for Lyapunov analysis after taking care of the transients
of the initial evolution. The proposed exploratory ranges for a chaotic
basin and finally retained values are given in Table.\ref{tab:Parameter-values-for}.
These ranges define a numerical search, not measured plasma parameters
or an established chaotic regime, but is indicative of the respective
physics as indicated in the table. 

\subsubsection{Exploratory parameter range}

As surrogate model equations are now normalized, the typical variables
satisfy $|\xi|\sim1$ and $|z|\sim1$. The Stuart-Landau regime is
derived for a weakly unstable mode with IICSI growth rate $\gamma_{\textrm{IICSI}}\ll\omega_{0}$.
So, naturally $0<\epsilon\ll1$. If the charge-relaxation parameter
$\Lambda\gg1$, the relaxation responds much faster than the dominant
mode while $\Lambda\ll1$ indicates that debris charge evolves very
slowly. This suggests the physics-guided range $0.1\lesssim\Lambda\lesssim1.0$.
The exploratory numerical scan is nevertheless extended to $\Lambda=10$
in Table.\ref{tab:Parameter-values-for} to test whether a faster
charging response creates additional branches.

To estimate oscillator-feedback parameters, we first consider the
instantaneous oscillator force balance at fixed charge. Setting $y=0$
and $y'=0$ in Eq.(\ref{eq:y}) gives the oscillator nullcline
\begin{equation}
\left(\omega^{2}_{0}+\beta q\right)\xi=\kappa q.
\end{equation}
This relation is not, by itself, the equilibrium condition for the
complete $(\xi,y,z)$ system. A full equilibrium must additionally
satisfy
\begin{equation}
y=0,\qquad(1+Hz)\xi=Kz,\qquad C\xi-\Lambda z+E\xi z-Gz^{3}=0.
\end{equation}
The following estimates for $H$ and $K$ are therefore conditional
nullcline estimates, intended only to prevent the charge feedback
from displacing the oscillator far outside the amplitude range of
the van der Pol surrogate. So, we define 
\begin{equation}
\omega^{2}_{\textrm{eff}}=\omega^{2}_{0}+\beta q.
\end{equation}
Using the normalized variables, we can write
\[
\omega^{2}_{\textrm{eff}}=\omega^{2}_{0}(1+Hz)
\]
following which we define a change in the squared frequency relative
to $q=0$ as
\[
\begin{aligned}\Delta\omega^{2}_{\textrm{eff}} & \equiv\omega^{2}_{\textrm{eff}}(q)-\omega^{2}_{0}=\beta q\end{aligned}
\]
so that, for $|z|\sim1$, the characteristic magnitude of $H$ can
be estimated from
\begin{equation}
H\sim\frac{\Delta\omega^{2}_{\textrm{eff}}}{\omega^{2}_{0}}.
\end{equation}
Naturally, the charge-induced frequency correction should not be excessively
large within the weak-feedback interpretation, so that $|Hz|\lesssim1$.
A sensible physics-guided range is therefore $|H|\sim0.01-0.3$; the
search is widened modestly to $|H|\leq0.5$ in Table.\ref{tab:Parameter-values-for}.
The coefficient $K$ measures the modal displacement force caused
by a characteristic charge fluctuation. On the oscillator nullcline,
from Eq.(\ref{eq:y}), we have 
\begin{equation}
\xi_{\textrm{eq}}=\frac{Kz}{1+Hz},
\end{equation}
provided $(1+Hz)\neq0$. Thus, along the nullcline, for $z\sim1$,
$|1+Hz|\sim1$, and $|\xi|\lesssim1$, we obtain the conditional estimate
$|K|\lesssim1$. 

In order to estimate $C$ and $D$, we note that the basic charging
equation, Eq.(\ref{eq:z}) can be approximated as
\begin{equation}
z'+\Lambda z=C\xi+D\xi'.
\end{equation}
Assuming the dominant IICSI frequency $\bar{\omega}$ to be approximately
harmonic, we can write
\begin{equation}
\xi(\tau)=\xi_{a}\cos(\bar{\omega}\tau),\qquad\xi'(\tau)=-\bar{\omega}\xi_{a}\sin(\bar{\omega}\tau),
\end{equation}
so that the forcing on the right-hand side of the charging equation
is
\begin{equation}
C\xi+D\xi'=C\xi_{a}\cos(\bar{\omega}\tau)-D\bar{\omega}\xi_{a}\sin(\bar{\omega}\tau)
\end{equation}
and the combined forcing amplitude is $\xi_{a}\sqrt{C^{2}+\bar{\omega}^{2}D^{2}}$.
The steady harmonic response of the left-hand side has amplitude $z_{a}\sqrt{\Lambda^{2}+\bar{\omega}^{2}}$.
Equating these amplitudes gives 
\begin{equation}
C^{2}+\bar{\omega}^{2}D^{2}=r^{2}\left(\Lambda^{2}+\bar{\omega}^{2}\right).
\end{equation}
where $r=z_{a}/\xi_{a}$ is the response ratio. So, the admissible
pairs lie on the ellipses defined by the above equation. However,
this amplitude relation does not give any clue to the sign of $C$
or $D$. Physically,
\begin{equation}
C\propto\sum_{X}\left(\frac{\partial I_{{\rm tot}}}{\partial X}\right)_{0}A_{X},\qquad D\propto\sum_{X}\left(\frac{\partial I_{{\rm tot}}}{\partial X}\right)_{0}B_{X},
\end{equation}
where $A_{X}$ and $B_{X}$ are respectively the in-phase and quadrature
projections of the plasma observable $X$ onto the IICSI mode. Depending
on whether that phase increases electron collection, ion collection,
photoelectron escape, SEE, or electron return, either $C>0$ or $C<0$
is physically admissible and the same is true of $D$. The coefficients
are nevertheless effective responses of the same plasma-surface system,
so an experimental change can alter several of their magnitudes simultaneously.
The feedback coefficients $H$ and $K$ may likewise have either sign,
because they describe the projected frequency shift and direct modal
force associated with a signed charge perturbation. By contrast, $\epsilon>0$
denotes the self-excited branch, $\Lambda>0$ is passive linear charge
relaxation, and $G>0$ makes the nonlinear saturation possible. The
branch investigated below uses $E<0$, for which $E\xi z$ can reduce
the instantaneous relaxation during one phase of the oscillation and
thereby provide nonlinear charging gain.

Only under the additional restriction $r\lesssim1$, we have the bounds
as
\begin{eqnarray}
|C| & \lesssim & \sqrt{\Lambda^{2}+\bar{\omega}^{2}},\\
|D| & \lesssim & \frac{1}{\bar{\omega}}\sqrt{\Lambda^{2}+\bar{\omega}^{2}}.
\end{eqnarray}
For $\bar{\omega}\simeq1$ and $\Lambda\lesssim1$, this restricted
bound estimate gives $|C|,|D|\lesssim1-1.4$. However, we should note
that it is not a universal bound. The numerical search is subsequently
widened to $[-3,3]$ because the normalization $Q_{0}$ does not require
$z_{a}\leq\xi_{a}$. For the retained values $C=-2.3761$, $D=-0.9267$,
$\Lambda=0.1907$, and $\bar{\omega}\simeq1$, the response relation
gives $r\simeq2.5$. Therefore the retained $C$ exceeds the restricted
$r\leq1$ bound but is consistent with the wider, explicitly stated
large-response search.

From Eq.(\ref{eq:z}), the effective charge-relaxation rate is given
by
\begin{equation}
\Lambda_{\mathrm{eff}}=\Lambda-E\xi+Gz^{2}.
\end{equation}
If the IICSI mode amplitude is $\xi_{a}$, the nonlinear modulation
has strength $|E|\xi_{a}$, then an useful exploratory range can be
\begin{equation}
0.1\lesssim\frac{|E|\xi_{a}}{\Lambda}\lesssim2.
\end{equation}
or in other words $|E|\sim\Lambda$ for $\xi_{a}\sim1$. We also note
that the cubic term in Eq.(\ref{eq:z}) becomes important when $|G|z^{2}_{a}\sim\Lambda,$
where $z_{a}$ is the amplitude of charge oscillation. For $z_{a}\sim1$,
we again have $|G|\sim\Lambda$.
\begin{table}
\caption{Lyapunov exponents in the chaotic basin.\vrdd}\label{tab:Lyapunov-exponents-in}

\centering{}%
\begin{tabular}{ccccccc}
\hline 
$\Delta\tau$ & \quad{} & $\lambda_{1}$ & \quad{} & $\lambda_{2}$ & \quad{} & $\lambda_{3}$\vrr\tabularnewline
\hline 
\hline 
$0.02$ &  & $0.07781$ &  & $-1.33\times10^{-4}$ &  & \vrtt$-1.2850$\vrd\tabularnewline
$0.01$ &  & $0.07959$ &  & $-8.91\times10^{-5}$ &  & $-1.2691$\vrd\tabularnewline
$0.005$ &  & $0.07803$ &  & $-5.87\times10^{-4}$ &  & $-1.2706$\tabularnewline
\hline 
\end{tabular}
\end{table}

Finally, we note that the parameters $\epsilon,\Lambda,$ and $G$
are fixed by physical construction of the theoretical model where
$\epsilon$ represents the linear growth of the self-excited IICSI
mode, $\Lambda$ represents passive linear charge relaxation, and
$G$ is the nonlinear saturation coefficient of the charging coordinate,
which is essential for bounded large-amplitude charge dynamics. Naturally,
we \emph{must} have all $(\epsilon,\Lambda,G)>0$. In contrast to
this, the signs of $C,D,H,K,$ and $E$ are not universally fixed
and they are determined by the local stability scenario.

\subsubsection{Analysis of the theoretical results}

Table.\ref{tab:Lyapunov-exponents-in} shows the Lyapunov exponents
of this dynamical evolution for initial conditions $(\xi,y,z)=(0.1,0,0)$.
The stable positive exponent $\lambda_{1}$, a near-zero second exponent
$\lambda_{2}$, and a negative third exponent $\lambda_{3}$ establish
a dissipative chaotic attractor on this branch. Other tested initial
conditions approach a regular attractor with $\lambda_{1}\simeq0$,
revealing multistability \cite{feudel2008}. So, the theoretical result
is an existence proof for an allowed chaotic basin, not global attraction
from every initial condition, which is also expected as these plasma
systems by themselves cannot be chaotic in all parameter regimes. 

\begin{figure}
\begin{centering}
\includegraphics[width=1\textwidth]{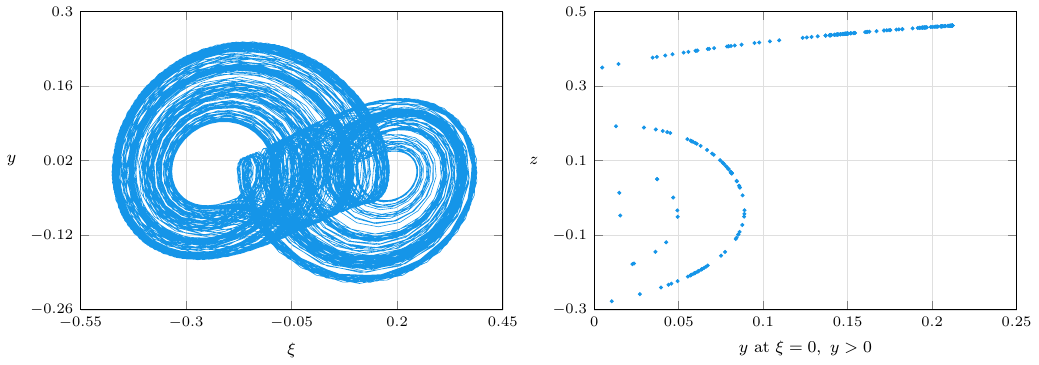}
\par\end{centering}
\caption{The folded phase-space $(\xi,z)$ (left) and the Poincaré section
for the chaotic basin.}\label{fig:The-folded-phase-space}
\end{figure}

\begin{figure}
\begin{centering}
\includegraphics[width=1\textwidth]{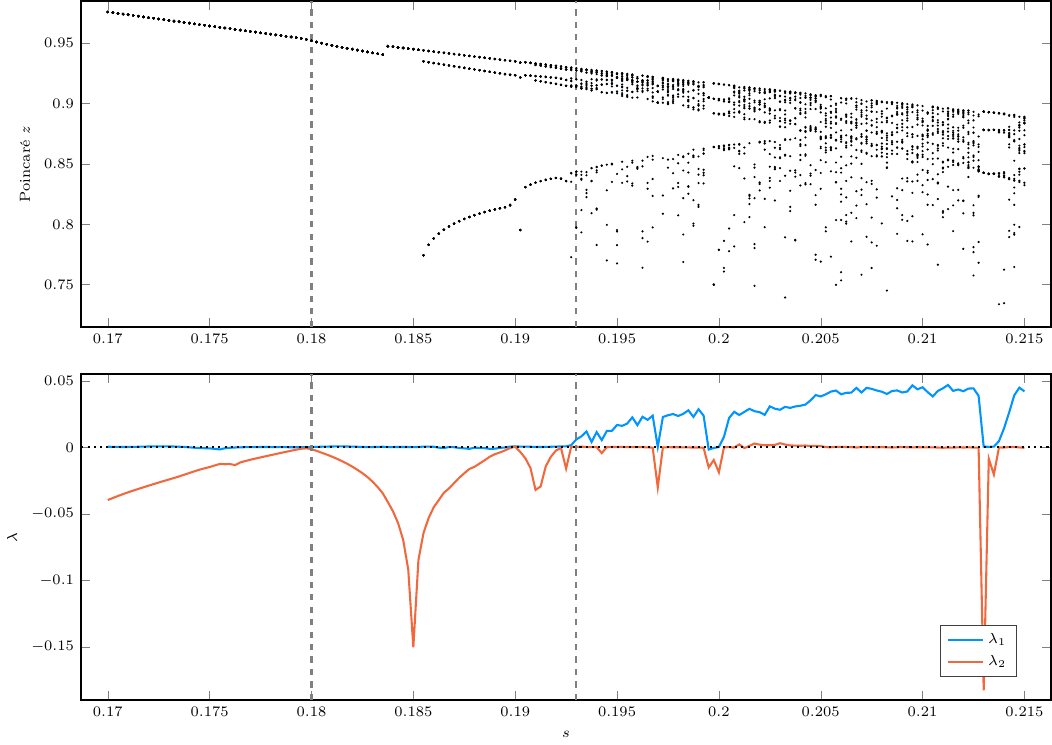}
\par\end{centering}
\caption{Poincaré splitting leading to the chaotic basin (upper panel). The
corresponding Lyapunov exponents. The two vertical dashed lines at
$s=0.18$ and $0.193$ indicate the shift in the dynamics as indicated
by the Lyapunov exponents.}\label{fig:Poincare-splitting-leading}
\end{figure}

Fig.\ref{fig:The-folded-phase-space} shows the folded phase-space
$(\xi,z)$ along with the Poincaré section of the chaotic attractor.
To explore the route to chaos, the evolution is continued with 
\begin{equation}
E(s)=sE_{*},\qquad G(s)=sG_{*},\qquad0\le s\le1,
\end{equation}
while $(\epsilon,\Lambda,H,K,C,D)$ were held constant at the chaotic
parameter set (see Table.\ref{tab:Parameter-values-for}). The values
$E_{*}$ and $G_{*}$ are the values indicated at Table.\ref{tab:Parameter-values-for},
around which the values are changed through $s$. The results of these
scans are shown in Fig.\ref{fig:Poincare-splitting-leading}. Each
trajectory is first evolved for a dimensionless transient interval
$\Delta\tau_{\mathrm{tr}}=1000$. With the time step $\Delta\tau=0.02$,
this corresponds to $5.0\times10^{4}$ fourth-order Runge-Kutta steps.
Poincaré crossings at $\xi=0$ with $y>0$ are then accumulated for
a further dimensionless interval $\Delta\tau=1800$ ($9.0\times10^{4}$
integration steps), and up to 80 crossings are plotted for each $s$.
A coarse 201-point scan is carried for $0\le s\le1$ and then a fine
181-point scan is covered for $0.17\le s\le0.215$ in increments of
$2.5\times10^{-4}$. The resulting Poincaré section has one point
through $s\simeq0.1795$, which then splits into two near $s=0.17975$,
and develops further finite-cycle splitting around $s\simeq0.1855-0.19$,
and broadens into multi-band / chaotic sections near $s\simeq0.193$.
The numerically resolved route is therefore dominated by period-doubling
bifurcations, followed by band formation and periodic windows, rather
than a simple torus-breakdown scenario.

\section{PIC implementation}

For the purpose of testing our reduced model, we employ our well-tested
1D \emph{hybrid}-PIC-MCC code with open-boundary configuration \cite{dawson1983,hockney1988,birdsallLangdon1991,birdsall1991,verboncoeur2005,changmai2019,changmai2020,dasChangmai2023}.
With the debris placed in the middle of the simulation box, the open-boundary
configuration ensures that the simulation can be run as long as required
without any boundary effects, which is necessary for long dynamic
evolution required for detecting chaotic basin, if any. The PIC observables
used as a candidate for $\xi$ are local ion mean velocity $\left\langle u_{i}\right\rangle $,
debris potential $\phi_{d}$, debris potential-well depth, local ion
and electron densities $n_{i,e}(x_{d})$ at debris position $x_{d}$.
The code dumps the Vlasov snapshots from which these observables can
be easily estimated.

For the Vlasov observable, ion distribution frames are binned on $64\times64$
phase-space cells every 25 time steps. Densities are estimated from
integrating over velocity from which their temporal mean is removed
and a singular value decomposition (SVD) is performed. The local history
is preserved which is later used for SINDy-type weak regression \cite{brunton2016,rudy2017,messenger2021}.

\subsection{Photoemission}

As discussed before, photoemission as well as SEE supplies the much
needed linear and nonlinear terms in the charging evolution equation
at the PIC level. Particularly, photoemission along with maintaining
a current bias also provides a population of low-energy emitted electrons
whose escape and return are sensitive to the local surface barrier.
It consequently contributes to the linear coefficients $\nu_{\mathrm{ch}}$,
$a$, and $b$, and helps place the system in the region in which
the closure is applicable. Along with SEE, photoemission helps establish
a regulated operating charge $Q_{d0}$ satisfying the relation
\begin{equation}
I_{\textrm{tot}}(Q_{d0})=(I_{e}+I_{i}+I_{\mathrm{ph}}+I_{\mathrm{SEE}})|_{Q_{d0}}=0.
\end{equation}
A representation of the photoemission current can be written as
\begin{equation}
I_{\mathrm{ph}}=I^{(0)}_{\mathrm{ph}}(\varPhi_{\gamma},{\cal M})\,P^{\mathrm{esc}}_{\mathrm{ph}}(Q_{d},\bm{X}),
\end{equation}
where $P^{\mathrm{esc}}_{\mathrm{ph}}$ is the probability of escape
of emitted photoelectrons, and $\varPhi_{\gamma}$ is the incident
photon flux, and ${\cal M}$ represents material properties of the
debris, which determines the photoelectron yield $Y_{\mathrm{ph}}$. 

\subsection{SEE with surface regulation}

For SEE, we can write the net SEE current as \cite{draine1979,chow1993,sternglass1950}
\begin{equation}
I_{\mathrm{SEE}}\simeq e\int\delta_{\mathrm{SEE}}(\mathcal{E},{\cal M})\,P^{\mathrm{esc}}_{\mathrm{SEE}}(Q_{d},\bm{X},{\cal E})\,\Gamma_{e}(\mathcal{E},t)\,d\mathcal{E},
\end{equation}
where $\delta_{\mathrm{SEE}}(\mathcal{E})$ is the energy-dependent
yield and $\Gamma_{e}$ is the incident electron flux for the impact
energy ${\cal E}$, and $P^{\mathrm{esc}}_{\mathrm{SEE}}$ is the
probability of escape of emitted secondary electrons. As the IICSI
perturbs density, potential, electron flux, and the energy distribution
arriving at the debris, the SEE current can change with $Q_{d}$,
\begin{equation}
\frac{\partial}{\partial Q_{d}}I_{\mathrm{SEE}}(Q_{d,}\bm{X})\simeq\left.\frac{\partial I_{\mathrm{SEE}}}{\partial Q_{d}}\right|_{0}+\sum_{X}\left.\frac{\partial^{2}I_{\mathrm{SEE}}}{\partial Q_{d}\partial X}\right|_{0}\delta X.\label{eq:see-slope}
\end{equation}
Multiplication by $q$ and projection onto the IICSI mode, as outlined
before, produce the following terms
\begin{equation}
q\text{\ensuremath{\sum_{X}\left.\frac{\partial^{2}I_{\mathrm{SEE}}}{\partial Q_{d}\partial X}\right|_{0}\delta}X}\simeq c\xi q+d\dot{\xi}q\simeq c\xi q.\label{eq:see-projection}
\end{equation}
While the photoemission provides the persistent emission bias and
an independently regulated low-energy population of electrons, SEE
supplies the strong IICSI-sensitive gain.

We now consider the fact that net photoemission and SEE current is
regulated by the debris surface potential i.e.\ the recollection
current out of the net photoemission and SEE currents. Let us consider
the perturbation in the surface potential of the debris with respect
to the reference state be $\delta\phi_{d}$. Expanding the total regulated
current, we get,
\begin{equation}
I_{\mathrm{reg}}(q,\bm{X}_{0})=I_{Q}q+\frac{1}{2}I_{QQ}q^{2}+\frac{1}{6}I_{QQQ}q^{3}+\cdots,\label{eq:regulated-expansion}
\end{equation}
which supplies the term $-gq^{3}$ in Eq.(\ref{eq:evol3}) when the
local third derivative is negative, $I_{QQQ}<0$, consistent with
$g=-I_{QQQ}/6>0$.

\subsection{Numerical implementation}

The \emph{h}-PIC-MCC code is a proven and benchmarked one-dimensional
electrostatic code which can handle external debris of arbitrary size
with photoemission and SEE \cite{hockney1988,birdsallLangdon1991,birdsall1991,verboncoeur2005,changmai2019,changmai2020,dasChangmai2023},
both with surface regulation in variable boundary configuration including
open-boundary. The open-boundary (see below) ensures that plasma can
continuously flow across the simulation domain with any arbitrary
velocity and the simulation can be run indefinitely without any boundary-induced
effects. This is essential, when we want run the simulation long enough
to see if there are chaotic components. The code can handle a domain
length of arbitrary length and given a particular plasma configuration
in a certain domain length, the code can automatically scale its numerical
parameters so that the plasma configuration remain same in a new domain
length. The primary simulation parameters of the code are given in
Table.\ref{tab:pic}.

\begin{table}
\begin{centering}
\caption{Numerical and plasma parameters.}\label{tab:pic}
\begin{tabular}{@{}l@{}ll@{}}
\toprule 
\textbf{Quantity} & $\quad$ & \textbf{Value}\tabularnewline
\midrule 
Time step &  & $\omega_{pe}\Delta t=0.05$\tabularnewline
Domain and grid &  & $L=350\lambda_{D}$, $N_{x}=1024$$^{\dag}$\tabularnewline
Boundary &  & Bounded, periodic, and open\tabularnewline
Boundary conditions &  & Dirichlet, Neumann, periodic, and mixed\tabularnewline
Reference particle numbers &  & $N_{e}=N_{i}=167000$$^{\dag}$\tabularnewline
Temperatures and mass ratio &  & $T_{e}=1$, $T_{i}=0.01$, $m_{i}/m_{e}=1836$\tabularnewline
Boundary potential &  & $\phi_{L}=\phi_{R}=0$\tabularnewline
Debris position and size &  & $x_{d}/L\in[0,1]$, size $\sim1\lambda_{D}$\tabularnewline
Charging &  & Dynamic\tabularnewline
Macro-particle physical weight &  & $10^{6}$ physical particles per macro-particle\tabularnewline
Plasma flow &  & Subsonic to supersonic\tabularnewline
\bottomrule
\end{tabular}
\par\end{centering}
\raggedright{}$^{\dag}$\emph{These are only reference. The code automatically
scales the parameters based on the domain length and given plasma
configuration.}
\end{table}

\subsubsection{Open-boundary reservoir, buffer, and regulated equilibrium}

The open-boundary uses reservoirs at the boundaries \cite{birdsallLangdon1991,verboncoeur2005,changmai2020,dasChangmai2023},
which removes outgoing particles and independently injects distribution-consistent
particles from both ends. A boundary buffer of length $20\lambda_{D}$
is used to for particle relaxation, so that the plasma is \emph{not}
affected by particle entering/leaving the domain. However, we note
that long runs can nevertheless accumulate a common electron-ion density
deficit (or surplus), which is why we use an optional regulator, which
continuously measures the common deficit/surplus relative to baseline
$n_{e}=n_{i}=1$ and restores charge-neutral electron-ion pairs. Its
response time is $100\omega^{-1}_{pe}$ and its correction is capped
at $10^{-3}$ of the target population per time step. The insertion
(or removal) of charge-neutral pair is essential for regulation of
the reservoir inventory without directly charging the domain or debris.
\begin{figure}
\begin{centering}
\includegraphics[width=1\textwidth]{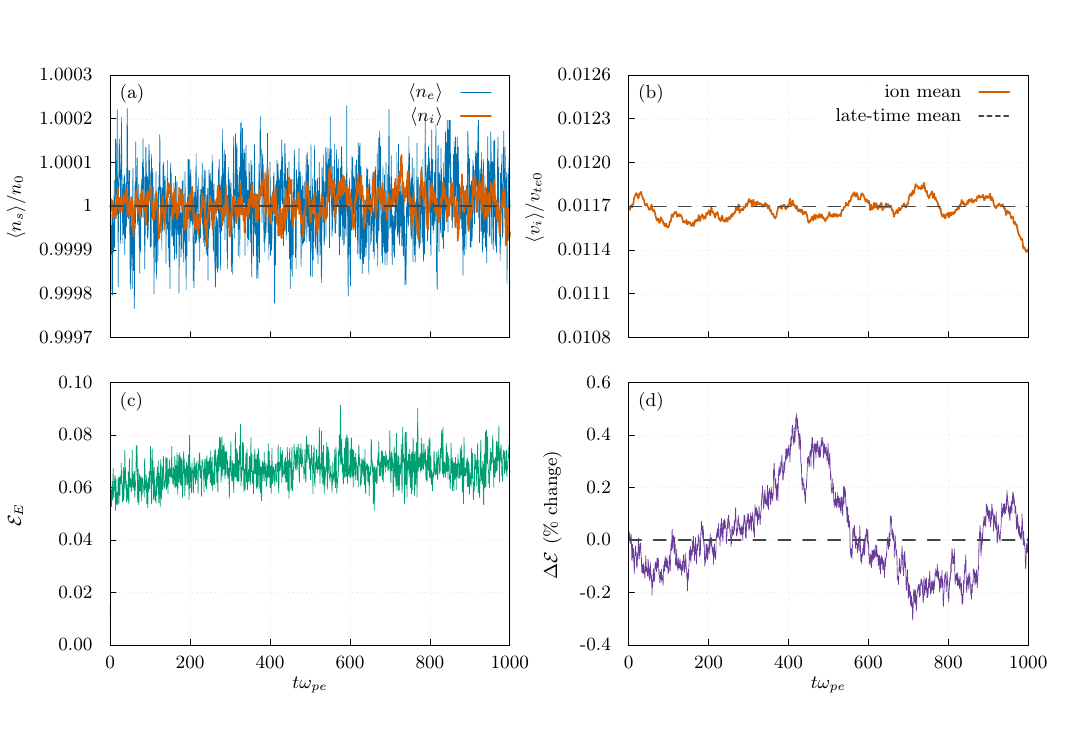}
\par\end{centering}
\caption{Diagnostic plots for open-boundary PIC simulation without any debris
(control run) and with equilibrium plasma subsonic flow $M=0.5$.
(a)\ Domain-averaged electron and ion densities $\left\langle n_{i,e}\right\rangle $,
normalized by the equilibrium density $n_{0}$. (b)\ Mean ion velocity.
The dashed line denotes its late-time mean. (c)\ Electrostatic field
energy ${\cal E}_{E}$. (d)\ Percentage change in the total simulated
energy $\Delta{\cal E}$ relative to its initial value.}\label{fig:Diagnostic-1}
\end{figure}

The charge on the external debris can be set to any value -- either
positive or negative (without an active bias) and thereafter, electron
and ion capture determines the charge $Q_{d}(t)$ stochastically.
We use a maximum electron capture probability of $0.020$ at energy
scale $0.20$ and ion capture probability of $0.010$ at $0.05$.
Both thresholds are zero in the normalized energy unit. The particle
capture algorithm uses Monte Carlo collision decision.
\begin{figure}
\begin{centering}
\includegraphics[width=1\textwidth]{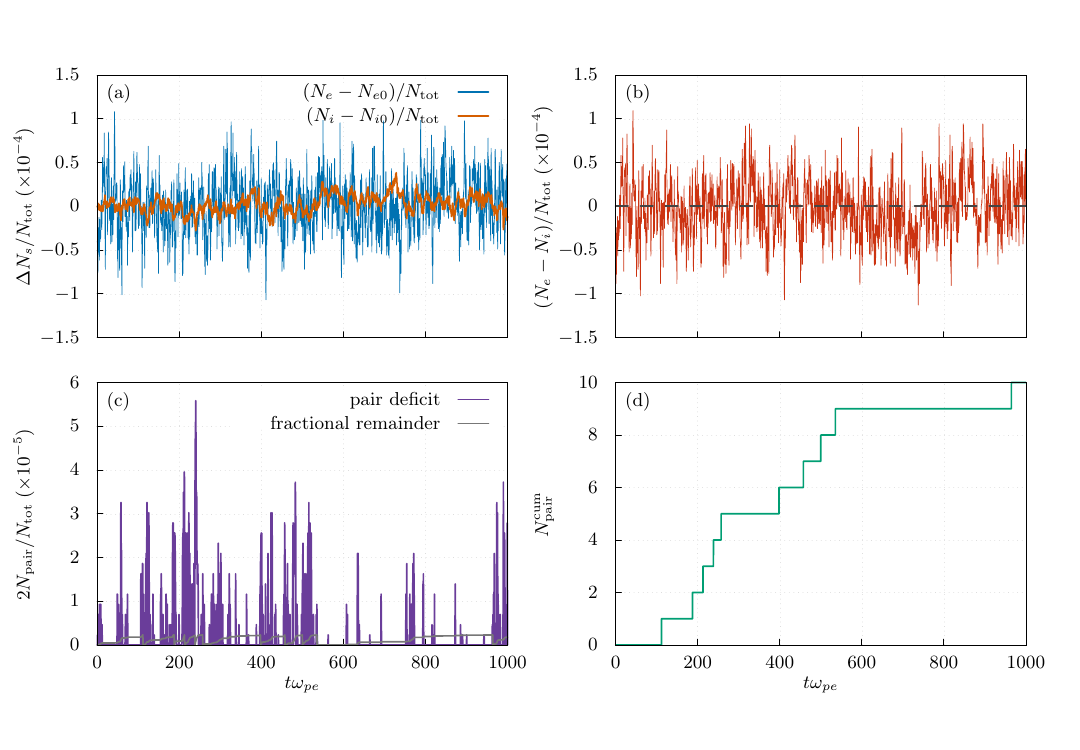}
\par\end{centering}
\caption{Particle-inventory and density-regulator diagnostics for the no-debris
control simulation. (a)\ Changes in the electron and ion macroparticle
populations relative to their initial values, expressed as fractions
of the initial total population. (b)\ Electron-ion population imbalance
$(N_{e}-N_{i})/N_{\mathrm{tot}}$. the dashed line denotes exact macroparticle
charge balance. (c)\ Charge-neutral pair deficit and the fractional
controller remainder, expressed as $2N_{\mathrm{pair}}/N_{\mathrm{tot}}$.
(d)\ Cumulative number of charge-neutral electron-ion pairs injected
by the reservoir-density regulator. Together, the panels show that
the regulator maintains the open-system particle inventory while introducing
only small, charge-neutral corrections (only about 10 pairs till $t=1000\omega^{-1}_{pe}$).}\label{fig:Diagnostic-2}
\end{figure}

The base PIC photoemitter uses a photon flux of $2.5\times10^{15}\,\mathrm{\mathrm{m}^{-2}s^{-1}}$,
with a photon energy of $20\,\mathrm{eV}$, quantum yield of $0.10$.
A directional illumination fraction $0.50$ is used. The reference
density and temperature for normalization are $n_{\mathrm{ref}}=10^{11}\,\mathrm{\mathrm{m}^{-3}}$
and $T_{e,\mathrm{ref}}=100\,\mathrm{\mathrm{eV}}$. Emission is a
cold/hot mixture with temperatures 2.7 and 10 eV and hot fraction
0.05. The emitting area is derived from the debris size. Each emitted
electron is inserted as a real PIC macro-particle just outside the
appropriate debris surface, which decreases the negative debris charge
consistently, and may escape or return to the debris. Therefore photoemission
supplies both a current and a self-consistent feedback path.

In Figs.\ref{fig:Diagnostic-1} and \ref{fig:Diagnostic-2}, we show
the diagnostics plots for the open-boundary PIC configuration. Both
figures uses the same plasma configuration that is used later for
the investigative case but without any external debris. So, the figures
demonstrate how well the open-boundary simulation mimic a flowing
plasma. Fig.\ref{fig:Diagnostic-1} shows the domain-averaged electron
and ion densities {[}panel (a){]}, mean ion velocity {[}panel (b){]},
the electrostatic field energy ${\cal E}_{E}${[}panel (c){]} for
a discrete grid
\begin{equation}
\mathcal{E}_{E}(t)\simeq\frac{\Delta x}{2}\sum^{N_{x}}_{j=1}E^{2}_{j}(t),
\end{equation}
and percentage change in the total simulated energy $\Delta{\cal E}$
relative to its initial value, where ${\cal E}(t)$ is the total energy
\begin{equation}
\mathcal{E}(t)=\frac{w_{e}m_{e}}{2}\sum^{N_{e}(t)}_{p=1}v^{2}_{e,p}(t)+\frac{w_{i}m_{i}}{2}\sum^{N_{i}(t)}_{p=1}v^{2}_{i,p}(t)+\frac{\Delta x}{2}\sum^{N_{x}}_{j=1}E^{\,2}_{j}(t).
\end{equation}
In the above expression, $p$ is the computational particle index
and $w_{s}$ is particle weight. These results prove that the simulation
preserves the physical properties of the plasma as the plasma flows
across the simulation window, where we can see that the normalised
electron and ions densities remain close to unity without secular
particle loss, accumulation, or charge separation. The same is reflected
in the ion velocity, electrostatic energy, total energy. Fig.\ref{fig:Diagnostic-2}
shows the macroparticle (computational particle) inventory, which
shows the efficiency of the reservoir-density regulator.

\begin{figure}
\begin{centering}
\includegraphics[width=1\textwidth]{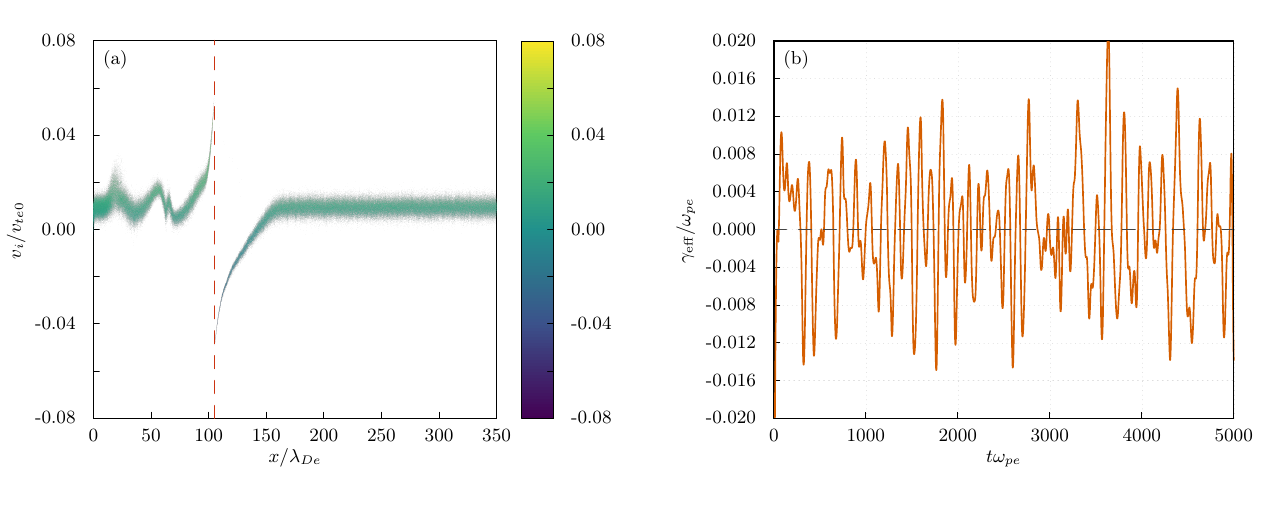}
\par\end{centering}
\caption{Ion phase-space {[}panel (a){]} and time-evolution of IICSI growth
(right) plasma-debris simulation with $M=0.5$ {[}panel (b){]}. Panel
(b) shows the effective growth rate $\gamma_{\mathrm{eff}}=d\ln A/dt$
($A$ being the amplitude of the mode) of the leading coherent ion-density
mode, evaluated from its band-limited envelope over $0.015\protect\leq\omega/\omega_{pe}\protect\leq0.05$.
Positive and negative values indicate instantaneous modal growth and
decay, respectively. The horizontal dashed line denotes $\gamma_{\mathrm{eff}}=0$.}\label{fig:Ion-phase-space-structure}
\end{figure}

Fig.\ref{fig:Ion-phase-space-structure} shows the ion phase-space
diagram (left panel) in the investigative case for plasma-debris interaction
which is being used for determining deterministic chaos. The time
evolution of growth rate $\gamma_{\textrm{eff}}$ of the leading order
IICSI mode is shown in the right panel of Fig.\ref{fig:Ion-phase-space-structure}.
The growth rate history shows that the IICSI mode successfully attains
a statistical equilibrium, which is required for chaos analysis.

\subsubsection{Surface regulation of photoelectrons}

The photoelectron sheath immediately surrounding the debris surface
is thinner than the PIC grid spacing. Its unresolved influence on
subsequent emission is therefore represented by a sub-grid surface-regulation
model. The resolved and unresolved parts of the calculation must be
distinguished carefully. Every photoelectron generated as per the
surface-regulated photoemission rate is inserted immediately into
the simulation as a PIC macro-particle and subsequently evolves under
the self-consistent electric field. The additional sub-grid state
used to regulate later emission is introduced below, after the potential
response has been defined. The surface-regulation model is activated
at $t\omega_{pe}=300$. Before this time the ordinary photoemitter
operates without the additional surface-regulation feedback. This
allows the initially imposed debris charge and the surrounding plasma
to establish a floating-potential baseline. After activation, rapid
potential fluctuations are separated from the slow evolution of the
mean debris potential by introducing the running baseline $\bar{\phi}_{d}$,
\begin{equation}
\frac{\mathrm{d}\bar{\phi}_{d}}{\mathrm{d}t}=\frac{\phi_{d}-\bar{\phi}_{d}}{\tau_{b}},\qquad\tau_{b}\omega_{pe}=100.
\end{equation}
Thus, $\bar{\phi}_{d}$ follows only slow changes in the debris potential
and $\Delta\phi=\phi_{d}-\bar{\phi}_{d}$ measures the instantaneous
deviation from the baseline. The effect of this deviation on photoelectron
escape and return is represented by the potential-response factor
\begin{equation}
g_{\phi}=\frac{2}{1+\exp\!\left(a_{\phi}\Delta\phi/\phi_{s}\right)},\qquad a_{\phi}=0.75,\qquad\phi_{s}=0.50.
\end{equation}
This factor varies continuously between approximately $0$ and $2$.
When the debris becomes more positive than its recent baseline, $\Delta\phi>0$,
the emitted electrons are more strongly attracted back toward the
surface. Accordingly, $g_{\phi}<1$ and the escape is reduced and
recollection is enhanced. Conversely, when the debris becomes more
negative relative to the baseline, $\Delta\phi<0\Rightarrow g_{\phi}>1$
and escape probability increases.

We now define the unresolved surface variable $N_{c}(t)$, which is
a single auxiliary state, measured in equivalent electron-number units,
that records the recent strength of the unresolved near-surface photoelectron
sheath. It is initialized to zero when the surface-regulation feedback
is activated. It represents the cumulative influence of photoelectrons
temporarily occupying the unresolved near-surface region, including
their subsequent escape and return. It is not a population of PIC
macro-particles. Its only dynamical role is to retain the finite-time
memory of recent emission and recollection which is needed to regulate
the rate at which subsequent photoelectrons are launched into the
resolved PIC domain. The continuous-time equation used for this purpose
is
\begin{equation}
\frac{\mathrm{d}N_{c}}{\mathrm{d}t}=f_{t}\dot{N}_{\mathrm{ph},\mathrm{launch}}-N_{c}\left(\frac{g_{\phi}}{\tau_{\mathrm{esc}}}+\frac{2-g_{\phi}}{\tau_{\mathrm{ret}}}\right),\label{eq:nc}
\end{equation}
where $\dot{N}_{\mathrm{ph},\mathrm{launch}}$ is the rate of physical
photoelectrons actually launched as resolved PIC particles, and $f_{t}$
is the fraction of that launch rate registered in the surface-memory
state. The two decay terms describe escape-like and return-like relaxation
of the unresolved sheath memory, with characteristic times $\tau_{\mathrm{esc}}$
and $\tau_{\mathrm{ret}}$, respectively. In the simulation, $f_{t}=0.5$,
$\tau_{\mathrm{esc}}\omega_{pe}=8$, and $\tau_{\mathrm{ret}}\omega_{pe}=4$.
These decay terms do not launch additional PIC electrons and do not
transfer charge to the debris. Their role is only to determine how
quickly the unresolved surface state forgets earlier emission. This
distinction prevents the same emitted electron from being counted
once as a resolved particle and again as a later release from $N_{c}$.

The normalized occupation of the surface-memory state is
\begin{equation}
\nu=\frac{N_{c}}{N_{\mathrm{cap}}},\qquad N_{\mathrm{cap}}=10^{7},
\end{equation}
and the associated space-charge suppression factor is given by
\begin{equation}
s_{c}=(1+\nu)^{-3/2}.\label{eq:s}
\end{equation}
Thus,
\begin{equation}
s_{c}\simeq1\quad\text{when}\quad N_{c}\ll N_{\mathrm{cap}},
\end{equation}
whereas $s_{c}$ decreases smoothly as the memory of recent near-surface
emission increases. The final multiplier coupling the unresolved state
to the physical photoelectron launch rate is
\begin{equation}
m_{\mathrm{esc}}=\max\!\left(0.10,g_{\phi}s_{c}\right).
\end{equation}
Here $g_{\phi}$ gives the instantaneous response to the surface-potential
barrier, while $s_{c}$ gives the finite-memory, space-charge suppression.
Their product multiplies the gross photoelectron-production rate.
The lower bound $0.10$ retains a small escaping high-energy component
and prevents irreversible numerical closure of the emission channel.
The exponent $3/2$ in Eq.(\ref{eq:s}) provides a smooth space-charge-limited
saturation rather than imposing a polynomial charging term directly.
For clarity, one numerical time step after activation is executed
in the following order:
\begin{enumerate}
\item The debris potential is sampled and $\bar{\phi}_{d}$, $\Delta\phi$,
and $g_{\phi}$ are updated.
\item The previously stored $N_{c}$ is advanced through the two decay terms
in Eq.(\ref{eq:nc}), before the new photoelectron source for the
present step is known. 
\item This decayed $N_{c}$ is used to calculate $\nu$, $s_{c}$, and $m_{\mathrm{esc}}$.
The gross physical photoemission rate is then multiplied by $m_{\mathrm{esc}}$. 
\item The accepted photoelectrons are created immediately as resolved PIC
macro-particles just outside the debris surface. The corresponding
emission charge is applied once to $Q_{d}$ by the ordinary photoemission
routine.
\item A fraction $f_{t}$ of the number just created is added once to $N_{c}$
as the discrete source contribution $f_{t}\dot{N}_{\mathrm{ph},\mathrm{launch}}$
in Eq.(\ref{eq:nc}). This addition does not create another electron
and does not apply another charge change. 
\item The resolved particles are then advanced by the ordinary PIC particle
pusher. If a photoelectron later strikes the debris, its collection
is handled once by the ordinary debris-collection rule.
\end{enumerate}
The decay update in step 2 and the source update in step 5 are the
operator-split numerical implementation of the single evolution equation
(\ref{eq:nc}). Thus the surface-regulation module of the code modifies
only the creation probability of subsequent photoelectrons -- it
neither replaces resolved photoelectron transport nor duplicates particle
or charge accounting. For comparison with the reduced theoretical
charging variable, a slow diagnostic coordinate is constructed as
\begin{equation}
Q_{\mathrm{eff}}=Q_{d}-eN_{c}.
\end{equation}
Here $Q_{d}$ is the physical signed debris charge used by the PIC
field solver and $N_{c}\geq0$ is the auxiliary surface-memory state.
Accordingly, $Q_{\mathrm{eff}}$ is a charge-like diagnostic proxy,
not a second conserved physical charge. It is used only when comparing
the simulation with the reduced slow charging variable; it is not
deposited on the grid, does not enter Poisson's equation, and does
not replace $Q_{d}$ in the particle or debris-charge update.

\subsubsection{Surface regulation for electron-impact SEE}

For SEE, we employ the Sternglass-type yield \cite{sternglass1950,chow1993},
\begin{equation}
\delta(\mathcal{E})=\min\left[\max\left(s_{y}\delta_{\max}xe^{1-x},0\right),\delta_{\mathrm{cap}}\right],\quad x=\frac{\mathcal{E}-\mathcal{E}_{\mathrm{th}}}{\mathcal{E}_{\mathrm{pk}}-\mathcal{E}_{\mathrm{th}}},
\end{equation}
for an incident electron of energy $\mathcal{E}$. The values used
are $s_{y}=0.10$ is the SEE yield scaling factor, $\delta_{\max}=1$
is the nominal maximum SEE yield, $\mathcal{E}_{\mathrm{pk}}=300\,\mathrm{\mathrm{eV}}$
is the impact energy at which the unscaled Sternglass curve reaches
its maximum, $\mathcal{E}_{\mathrm{th}}=10\,\mathrm{\mathrm{eV}}$
is the threshold energy for secondary emission, $\delta_{\mathrm{cap}}=1.5$
is upper yield limit imposed on the numerical model, and secondary
electron temperature of $2\,\mathrm{eV}$.

\subsubsection{Construction of theoretical variables $(\xi,y,z)$}

For a PIC observable, say $u(t)$, we first construct the quantity
$\left(u(t)-\widehat{u}_{\mathrm{trend}}(t)\right)$ by removing any
kind of local offset and linear drift $\widehat{u}_{\mathrm{trend}}(t)$,
it might have with
\begin{equation}
\widehat{u}_{\mathrm{trend}}(t)=a_{0}+a_{1}t,
\end{equation}
where
\begin{equation}
a_{1}=\frac{\sum^{N}_{i=1}(t_{i}-\bar{t})\,[u(t_{i})-\bar{u}]}{\sum^{N}_{i=1}(t_{i}-\bar{t})^{2}},\qquad a_{0}=\bar{u}-a_{1}\bar{t}.
\end{equation}
In the above relations,
\begin{equation}
\bar{t}=\frac{1}{N}\sum^{N}_{i=1}t_{i},\qquad\bar{u}=\frac{1}{N}\sum^{N}_{i=1}u(t_{i}).
\end{equation}
So,
\begin{equation}
\widehat{u}_{\mathrm{trend}}(t)\equiv\widehat{u}_{\mathrm{trend}}(t;u)=\left(\bar{u}-a_{1}\bar{t}\right)+a_{1}t.
\end{equation}
The dominant frequency $f_{0}$ of the IICSI mode is then selected
inside within an window $0.015\le f/\omega_{pe}\le0.05$. We use a
band-pass filter ${\cal B}$, centered on $f_{0}$, with half-width
$\max(0.35f_{0},2\Delta f)$, from which the theoretical variable
$\xi$ is constructed
\begin{equation}
\xi=\frac{u_{{\cal B}}(t)}{\sqrt{2}\,\sigma_{\mathcal{B}}},\qquad u_{{\cal B}}(t)=\mathcal{B}[u-\widehat{u}_{{\rm trend}}],
\end{equation}
where $\sigma_{{\cal B}}$ is the standard deviation
\begin{equation}
\sigma_{\mathcal{B}}=\sqrt{\frac{1}{N-1}\sum^{N}_{j=1}\left[u_{{\cal B}}(t_{j})-\overline{u_{{\cal B}}}\right]^{2}}=\sqrt{\frac{1}{N-1}\sum^{N}_{j=1}u^{2}_{{\cal B}}(t_{j})}
\end{equation}
 as $\overline{u_{{\cal B}}}\simeq0$. The quadrature is then computed
spectrally to get the other variable $y$, 
\[
y=\frac{1}{\omega_{0}}\frac{d\xi}{dt},\qquad\omega_{0}=2\pi f_{0}.
\]
The effective debris charge is de-trended and low-pass filtered below
$0.10\omega_{pe}$, and normalized by the base value $Q_{0}$, to
get the third variable 
\begin{equation}
z=\frac{\mathcal{L}[Q_{{\rm eff}}/e-\widehat{Q}_{{\rm trend}}/e]}{Q_{0}},
\end{equation}
where ${\cal L}$ is low-pass filter operator. In these simulations,
we have used $Q_{0}=10^{8}$.

\subsection{Methodological approach}

The analysis of the simulation run is carried out in two stages --
SINDy-type weak regression and chaos analysis \cite{brunton2016,rudy2017,messenger2021}.
While the weak regression tries to fit the coefficients of the theoretical
model to the PIC observables, the second stage carries out the chaos
analysis on actual PIC data. Thus, the chaos analysis does not depend
on the coefficients obtained from weak regression. In a nutshell,
weak regression shows whether the PIC simulation can capture the essential
physics described by the theoretical model and chaos analysis tests
whether the PIC simulation contains any chaotic component. Together,
they provide the much-needed proof-of-principle chaos route central
to the theme of this paper. In the chaos analysis, we carry out the
usual 0-1 test, Wolf reconstruction of the phase space trajectories
to determine the Lyapunov exponents \cite{takens1981,wolf1985,gottwald2004}
and entropy statistics. In the following subsections, we describe
each of these methods briefly and also present the results.

During all the subsequent analysis and test, we use finite-time windows
of the simulation instead of a one-time long stride of the simulation.
An open kinetic system is not expected to be represented by one stationary
three-variable closure over its entire history. Charging equilibration,
reservoir replenishment, ion transit and trapping, sheath rearrangement,
and slow changes in modal content can make the coefficients of Eqs.(\ref{eq:xi}-\ref{eq:z})
inferred from a global fit into averages of dynamically different
regimes. Such averaging may suppress a short interval in which the
IICSI mode and the charging response form a coherent feedback loop.
Therefore, we deliberately opt for a finite-time window analysis and
pay our attention to the local dynamics within that finite interval.
In any case, we do not expect a single three-variable closure to resolve
the global attractor of the full kinetic plasma. The whole analysis
i.e.\ weak regression as well as chaos test are carried out for two
plasma flows (across the debris) -- subsonic $(M=0.5)$ and supersonic
$(M=1.25)$, where $M=u_{0}/c_{s}$ is the ratio of equilibrium plasma
flow velocity and ion-acoustic speed.

\subsubsection{Finite-time-window analysis}

The finite-time analysis is performed as follows. The initial trend
removal, frequency selection, band-pass and low-pass filters, amplitude
scales, and construction of $(\xi,y,z)$ are carried out once on the
full simulation run. We then carry out a piecewise comparison for
the simulation window $600\le t\omega_{pe}\le2400$, in five adjacent
windows of width $360\omega^{-1}_{pe}$: $600-960$, $960-1320$,
$1320-1680$, $1680-2040$, and $2040-2400$. The complementary sliding
search uses window widths $W\omega_{pe}=240,300,360,$ and $480$,
with starting times advanced by $30\omega^{-1}_{pe}$ from $t\omega_{pe}=800$
to $1900$, retaining only windows with usable record.

\begin{figure}
\begin{centering}
\includegraphics[width=1\textwidth]{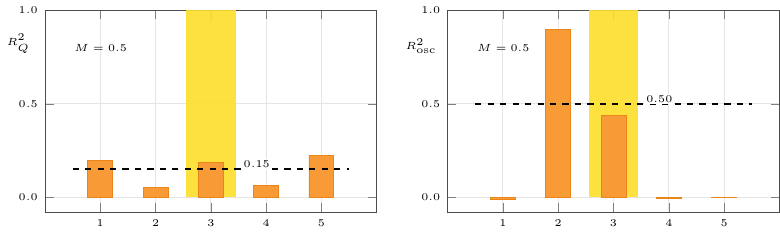}\\
\includegraphics[width=1\textwidth]{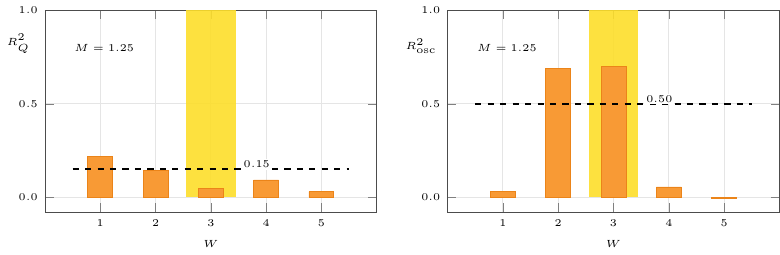}
\par\end{centering}
\caption{Sign-constrained PIC fitting results for the five fixed windows $W\omega_{pe}$
for $M=0.5$ (upper row) and $M=1.25$ (bottom row). The dashed lines
mark the thresholds $R^{2}_{Q}=0.15$ and $R^{2}_{\mathrm{osc}}=0.50$.
The shading identifies fixed window 3, $1320\protect\leq t\omega_{pe}\protect\leq1680$,
selected for the fixed-window comparison. The separate sliding-window
$M=0.5$ result at $1070\protect\leq t\omega_{pe}\protect\leq1550$
is reported in the text and is not one of these five fixed windows.}\label{fig:PIC-coefficient-fitting}
\end{figure}

\subsubsection{Weak regression}

To fit the coefficients appearing in the theoretical model, Eqs.(\ref{eq:xi}--\ref{eq:z}),
namely $(C,D,\Lambda,E,G)$ and $(\epsilon,H,K)$, we use a SINDy-type
weak (integral-form) regression \cite{brunton2016,rudy2017,messenger2021}.

Toward this, we use the following compact polynomial test function
\begin{equation}
w(s)=(1-s^{2})^{2},\qquad w'(s)=-4s(1-s^{2}),\qquad|s|\le1.
\end{equation}
The derivative is transferred to $w$ by integration by parts. Test
half-widths of 5, 10, and $15\omega^{-1}_{pe}$ are used, with the
center stride one third of the half-width. For the charging equation,
Eq.(\ref{eq:z}), each weak (integral-form) row contains 
\begin{equation}
\left[\int w\xi\,d\tau,\int wy\,d\tau,-\int wz\,d\tau,\int w\xi z\,d\tau,-\int wz^{3}\,d\tau\right]
\end{equation}
and target $-\int w'z\,d\tau$. For the oscillator equations, Eqs.(\ref{eq:xi} and \ref{eq:y}),
each row contains 
\begin{equation}
\left[\int w(1-\xi^{2})y\,d\tau,-\int w\xi z\,d\tau,\int wz\,d\tau\right]
\end{equation}
and target $-\int w'y\,d\tau+\int w\xi\,d\tau$. 

\subsubsection{Chaos analysis of the PIC projections}

The diagnostics required for chaos analysis of the PIC simulation
data are applied directly to the PIC observables. For this purpose,
each scalar record is smoothed with a 20-sample moving average and
sampled every 20 original steps. Since $\Delta t\omega_{pe}=0.05$,
the resulting diagnostic interval is $\Delta t_{{\rm diag}}\omega_{pe}=1$.
A de-trending for removal of the linear trend is carried out where
required. Charge level and charge increments are analyzed separately
as a stochastically accumulated charge can mimic diffusion in the
chaos 0-1 test and provide a false positive Lyapunov exponent.

The principal observables are $Q_{d}/e$, $\phi_{d}$, field energy,
and potential-well depth. Spectral entropy, peak fraction, autocorrelation
time, permutation entropy, nearest-neighbor forecast error, the 0-1
statistics, and determination of a Wolf-style largest Lyapunov exponent
are carried out. The Wolf-type largest Lyapunov exponent is evaluated
directly from the PIC-derived time series and compared with an ensemble
of phase-randomized reference signals. These reference signals preserve
the power spectrum and approximately preserve the linear autocorrelation
of the PIC record \cite{theiler1992,schreiber2000}, while destroying
nonlinear phase organization. They are therefore used only to construct
a statistical null comparison. A positive Wolf-type largest Lyapunov
exponent that exceeds the phase-randomized reference distribution
provides evidence of chaotic dynamics in the PIC trajectory that cannot
be explained by the linear spectral structure represented by that
reference ensemble alone. It does not, by itself, exclude every possible
nonlinear stochastic process.

\subsection{Results and interpretation}

\subsubsection{Reduced-model fitting to PIC data}

The post-run analysis train/test fit uses only the coefficients $(C,D,\Lambda)$
in the charge equation and $(\epsilon,H,K)$ in the oscillator equations.
The extended piecewise analysis includes $E$ and $G$ and is the
relevant structural test. Earlier exploratory fits enforced only $\Lambda,G,\epsilon\ge0$,
whereas the fits quoted below impose the complete sign pattern of
the retained theoretical branch. A ridge parameter $10^{-8}$ is used
in the piecewise analysis, and each standardized coefficient is required
to lie within a margin of $\eta=0.03$ inside its prescribed sign
domain. The coefficients of determination $R^{2}_{Q}$ and $R^{2}_{\mathrm{osc}}$
corresponding to the charging and oscillator equations for the best-fit
case are found to be 
\begin{equation}
\left(R^{2}_{Q},R^{2}_{\mathrm{osc}}\right)_{M=0.5}=(0.1611,0.5963),
\end{equation}
for the subsonic case $M=0.5$, in the common fitting interval $1070\leq t\omega_{pe}\leq1550$.
These values are above the corresponding threshold values of $0.15$
and $0.50$ for the charging and oscillation equations. The values
of the corresponding PIC fitting coefficients are
\begin{align}
(C,D,\Lambda,E,G)_{M=0.5} & =(-0.002884,-0.093300,0.059128,-0.140372,0.004507),\\
(\epsilon,H,K)_{M=0.5} & =(0.035615,0.861482,-0.048972),
\end{align}
for $M=0.5$, which conform to all the signs of the corresponding
coefficients of the theoretical model. To define the standardized
margin $\eta$, we write the weak regression as $\bm{b}=\bm{\varTheta}\bm{\theta}+\bm{r}$,
where $\bm{\varTheta}_{j}$ is the $j$th library column and $\bm{b}$
is the corresponding weak target. The dimensionless standardized coefficient
is 
\begin{equation}
\widetilde{\theta}_{j}=\frac{\|\bm{\varTheta}_{j}\|_{2}}{\|\bm{b}\|_{2}}\theta_{j}.
\end{equation}
If $s_{j}=+1$ or $-1$ denotes the sign required by the retained
theoretical branch, the standardized sign margin $\eta$ is imposed
through 
\begin{equation}
s_{j}\widetilde{\theta}_{j}\geq\eta.
\end{equation}
Thus, $\eta$ is a numerical sign-and-nonzero constraint rather than
a physical coefficient or a confidence level. It prevents a required
term from satisfying its sign constraint merely by collapsing to zero.
For the supersonic case $M=1.25$, the best sign-constrained fixed-window
PIC coefficients, obtained for $1320\leq t\omega_{pe}\leq1680$, are
\begin{align}
(C,D,\Lambda,E,G)_{M=1.25} & =(-0.002841,-0.044230,0.003886,-0.105340,0.024055),\\
(\epsilon,H,K)_{M=1.25} & =(0.003438,0.819830,-0.045350),
\end{align}
with
\begin{equation}
\left(R^{2}_{Q},R^{2}_{\mathrm{osc}}\right)_{M=1.25}=(0.0451,0.6991).
\end{equation}
So, the fitting fails the threshold for the charging equation for
the supersonic case.

As we can see, for the subsonic $M=0.5$ case, the PIC simulation
provides structural support for the theoretical charging-feedback
model within the favorable finite-time window which satisfies the
adopted goodness-of-fit thresholds for both the charging and oscillator
equations. This sign conformity of the fitted PIC coefficients with
the theoretical coefficients is important because it preserves the
physical direction of every feedback process -- modal driving of
charge, charge relaxation and nonlinear saturation, self-excitation
of the plasma mode, and the return coupling from charge to the oscillator.
We should however note that exact numerical agreement between the
PIC-fitted and theoretical coefficients is not required, because their
magnitudes depend on the normalization of the modal and charging coordinates,
the selected finite-time window, and the projection of a high-dimensional
kinetic plasma onto a three-variable reduced model. The theoretical
coefficients identify a chaotic regime of the reduced system, whereas
the PIC coefficients are effective local parameters which are affected
by unresolved modes, sheath kinetics, particle noise and reservoir
dynamics. The supersonic $M=1.25$ case is the weaker one because,
although the oscillator fit satisfactory, $R^{2}_{\mathrm{osc}}=0.6991$,
the charging fit with $R^{2}_{Q}=0.0451$ remains below the threshold
value of $R^{2}_{Q}=0.15$. Thus, the supersonic run reproduces part
of the oscillatory structure but not the simultaneous charge-oscillator
closure achieved for $M=0.5$, plausibly because faster flow introduces
stronger transit, trapping and multimode kinetic effects that cannot
be represented adequately by a single-mode, three-variable surrogate.

\subsubsection{PIC evidence for chaotic dynamics}

The chaos 0-1 test is applied to each de-trended and standardized
PIC observable $x_{j}$. An auxiliary angular parameter $c\in[0.37,2.77]$
is used to project the measured time series $x_{j}$ onto the two
translation variables $p_{n}$ and $q_{n}$,
\begin{equation}
p_{n}(c)=\sum^{n}_{j=1}x_{j}\cos(jc),\qquad q_{n}(c)=\sum^{n}_{j=1}x_{j}\sin(jc).
\end{equation}
These variables remain spatially bounded when the underlying dynamics
are regular, whereas chaotic or strongly diffusive fluctuations produce
an approximately unbounded random-walk-like trajectory in the $(p,q)$
plane. This behavior is checked by calculating the mean-square displacement
$M_{c}(n)$, evaluated for lags $n\leq\min(N/10,500)$. The correlation
between $M_{c}(n)$ and $n$ defines $K_{c}$,
\begin{equation}
K_{c}=\operatorname{corr}\!\bigl(n,M_{c}(n)\bigr).
\end{equation}
A value of $K_{c}\simeq0$ indicates that $\ensuremath{M_{c}(n)}$
does not grow systematically, indicating regular or quasi-periodic
dynamics, while a value of $K_{c}\simeq1$ indicates that $M_{c}(n)$
grows approximately with $n$, indicating chaotic or diffusive behavior.
Because an individual $c$ can accidentally resonate with a periodic
component of the signal, the test is performed for several $c$ values.
The final statistic is given by
\begin{equation}
K=\operatorname{median}_{c}\left(K_{c}\right).
\end{equation}
Because stochastic fluctuations can also generate diffusive translation
variables, a large $\ensuremath{K}$ in a set of noisy PIC data is
treated as an initial indication of irregular dynamics rather than
as independent proof of deterministic chaos. For the favorable $M=0.5$
subsonic interval, $1070\leq t\omega_{pe}\leq1550$, the 0-1 test
gives
\begin{equation}
K_{Q}=0.642,\qquad K_{\phi}=0.486,\qquad K_{\mathrm{field}}=0.595,\qquad K_{\mathrm{well}}=0.952.
\end{equation}
The simultaneous intermediate-to-large values found in several physically
related observables indicate that the irregular behavior is not confined
to the debris charge alone. In particular, the large value for the
potential-well observable suggests highly irregular evolution of the
self-consistent electrostatic structure. The 0-1 test on the successive
increments of the debris-charge time series, rather than from the
charge $Q_{j}$ itself, gives a $K_{\Delta Q}=0.998$, though the
nonlinear forecast error is statistically indistinguishable from that
of the phase-randomized reference ensemble. The near-unity value of
$K_{\Delta Q}$ is therefore attributed primarily to stochastic discrete-particle
or charging noise, rather than to deterministic chaos.

For the supersonic $M=1.25$ interval, $960\leq t\omega_{pe}\leq1680$,
the corresponding values are
\begin{equation}
K_{Q}=0.977,\qquad K_{\phi}=0.354,\qquad K_{\mathrm{field}}=0.520,\qquad K_{\mathrm{well}}=0.651,
\end{equation}
with $K_{\Delta Q}=0.999$. Although the debris-charge record has
a value close to unity, the weaker and less consistent response of
the potential, field and potential-well observables indicates that
the apparent diffusion is not coherently expressed across the coupled
plasma system. Moreover, the almost identical behavior of the charge
increments again points to a substantial stochastic contribution.
The large $\ensuremath{K_{Q}}$ value for $M=1.25$ therefore cannot,
by itself, be interpreted as evidence of deterministic chaos. In contrast,
the $M=0.5$ interval provides the more physically coherent candidate,
but again the chaos 0-1 test result \emph{must} still be considered
together with nonlinear predictability and Wolf-type Lyapunov exponent-based
tests.

To test whether the irregular PIC records retain nonlinear deterministic
organization, we also carry out a local nonlinear forecasting test.
For each de-trended and standardized observable $x_{j}$, a four-dimensional
delay embedding is constructed and the nearest admissible neighbor
is used to predict the signal ten diagnostic samples ahead. The normalized
root-mean-square prediction error is defined by \cite{farmer1987,takens1981}
\begin{equation}
{\rm NRMSE}=\left[\frac{\sum_{j}\left(x_{j+10}-\widehat{x}_{j+10}\right)^{2}}{\sum_{j}\left(x_{j+10}-\overline{x}\right)^{2}}\right]^{1/2},
\end{equation}
where $\widehat{x}_{j+10}$ is the nearest-neighbor prediction and
$\overline{x}$ is the mean of the predicted portion of the record.
The same calculation is repeated for eight PR reference signals. Since
these references preserve the power spectrum but randomize the Fourier
phases, a smaller PIC prediction error indicates temporal organization
that is not explained by the linear spectrum alone. For $M=0.5$ over
$1220\leq t\omega_{pe}\leq1460$, the PIC prediction errors for debris
potential, field energy, potential-well depth, and debris charge are,
respectively, 
\begin{equation}
1.075,\qquad0.895,\qquad1.211,\qquad1.356,
\end{equation}
whereas the corresponding PR-reference means and standard deviations
are 
\begin{equation}
1.302\pm0.061,\qquad1.071\pm0.081,\qquad1.331\pm0.129,\qquad1.345\pm0.121.
\end{equation}
The debris potential and field energy have lower prediction errors
than their phase-randomized (PR) references
\begin{equation}
1.075<1.302\pm0.061,\qquad0.895<1.071\pm0.081.
\end{equation}
This means that their temporal evolution contains predictive structure
that is destroyed when the Fourier phases are randomized. These collective
observables therefore carry evidence of nonlinear deterministic organization.

For $M=1.25$ over $960\leq t\omega_{pe}\leq1680$, the corresponding
PIC errors are 
\begin{equation}
1.208,\qquad0.558,\qquad0.632,\qquad1.083,
\end{equation}
compared with PR-reference values 
\begin{equation}
1.403\pm0.030,\qquad0.736\pm0.024,\qquad0.770\pm0.047,\qquad1.137\pm0.072.
\end{equation}
The supersonic run therefore also contains nonlinear deterministic
organization. This result is not, by itself, evidence of chaos, because
periodic and quasi-periodic dynamics can likewise be predictable.
The distinction is supplied by the independent Wolf comparison below
-- the $M=0.5$ case shows the strongest PR-relative Lyapunov excess,
whereas the $M=1.25$ case does not show a comparable separation.

For a Wolf-type largest Lyapunov exponent calculation \cite{wolf1985,rosenstein1993,kantz1994},
a delay embedding $\bm{X}_{i}=(x_{i},x_{i-\tau},\ldots,x_{i-(m-1)\tau})$
is constructed for $m=3,4,5$ and delays 20, 40, and 60 diagnostic
samples. Initial neighbors are excluded within a Theiler window $\max(100,2\tau)$.
The pair evolves until its separation reaches $d_{\max}=0.25,0.50,$
or 0.75, when a replacement neighbor is selected within a $60^{\circ}$
direction cone where possible. For every admissible parameter triplet,
the Lyapunov exponent is calculated through the relation 
\begin{equation}
\lambda_{\max}=\frac{\sum_{r}\log(d_{r}/d_{0,r})}{T},
\end{equation}
where $d_{0,r}$ is the initial separation at the beginning of the
$r$-th tracking interval. It is the distance between the reference
trajectory point and a nearby comparison point selected from a different
part of the reconstructed trajectory
\begin{equation}
d_{0,r}=\left\Vert \mathbf{X}(t_{r})-\mathbf{X}_{\mathrm{near}}(t_{r})\right\Vert 
\end{equation}
and $d_{r}$ is the evolved separation of that pair at the end of
the same tracking interval, immediately before the comparison point
is replaced or the separation is renormalized
\begin{equation}
d_{r}=\left\Vert \mathbf{X}(t_{r}+\Delta t_{r})-\mathbf{X}_{\mathrm{near}}(t_{r}+\Delta t_{r})\right\Vert .
\end{equation}
Here $T=\sum_{r}\Delta t_{r}$ is the total time over which the neighbor
pairs are followed. The reported statistic is the median over valid
embeddings, requiring at least five replacements. Thirty-two phase-randomized
reference records are evaluated with the same grid. The results of
this analysis is presented in Table.\ref{tab:Summary-of-results},
where the median Wolf-type largest Lyapunov exponent $\lambda^{\mathrm{PIC}}_{\mathrm{max}}$
is listed, which is obtained directly from the PIC-derived time series
over the tested reconstruction settings. Mathematically
\begin{equation}
\lambda^{\mathrm{PIC}}_{\max}=\operatorname{median}_{m,\tau,d_{\max}}\left[\lambda_{\max}(m,\tau,d_{\max})\right].
\end{equation}
For the $s$th phase-randomized (PR) reference signal, if $\lambda^{\mathrm{PR},s}_{\max}$
denotes the same median over accepted reconstruction settings, The
quantity $\overline{\lambda}^{\mathrm{PR}}_{\max}\pm\sigma^{\mathrm{PR}}_{\lambda}$
denotes the mean and standard deviation of the median maximum Lyapunov
exponents across an ensemble of $N_{\mathrm{PR}}$ phase-randomized
(PR) reference signals. For the $s$th PR signal, let $\lambda^{\mathrm{PR},s}_{\max}$
denote the median exponent obtained over all accepted reconstruction
settings. The ensemble mean is then given by
\begin{equation}
\overline{\lambda}^{\mathrm{PR}}_{\max}=\frac{1}{N_{\mathrm{PR}}}\sum^{N_{\mathrm{PR}}}_{s=1}\text{\ensuremath{\overline{\lambda}^{\mathrm{PR},s}_{\max}}},
\end{equation}
with the corresponding standard deviation
\begin{equation}
\sigma^{\mathrm{PR}}_{\lambda}=\sqrt{\frac{1}{N_{\mathrm{PR}}-1}\sum^{N_{\mathrm{PR}}}_{s=1}\left(\text{\ensuremath{\overline{\lambda}^{\mathrm{PR},s}_{\max}}}-\overline{\lambda}^{\mathrm{PR}}_{\max}\right)^{2}}.
\end{equation}

\begin{table}
\centering{}\caption{Summary of results from Wolf-type Lyapunov analysis.}\label{tab:Summary-of-results}
\begin{tabular}{lrrcc}
\toprule 
Case/window & $\lambda^{\mathrm{PIC}}_{\mathrm{max}}$ & $\overline{\lambda}^{\mathrm{PR}}_{\max}\pm\sigma^{\mathrm{PR}}_{\lambda}$ & $p$ & Chaos evidence\tabularnewline
\midrule 
$M=.5$, 1070--1550 & .2895 & $.2155\pm.0340$ & .061 & weak finite-time chaos evidence\tabularnewline
$M=.5$, 1220--1460 & .3120 & $.2614\pm.0786$ & .152 & positive, but not separated\tabularnewline
$M=.5$, 600--960 & .0465 & $.0435\pm.0094$ & $>0.05$ & no chaos evidence\tabularnewline
$M=.5$, 1680--2040 & .3567 & $.4289\pm.0395$ & -- & below reference\tabularnewline
$M=1.25$, 960--1320 & .2918 & $.3260\pm.0480$ & -- & below reference\tabularnewline
$M=1.25$, 1320--1680 & .2622 & $.2909\pm.0548$ & -- & below reference\tabularnewline
$M=1.25$, 960--1680 & .4501 & $.4469\pm.0361$ & $>0.05$ & indistinguishable\tabularnewline
$M=1.25$, 600--2400 & .2613 & $.2287\pm.0212$ & .121 & weak global excess\tabularnewline
\bottomrule
\end{tabular}
\end{table}
We also list the quantity $p$, the empirical probability under the
phase-randomized null hypothesis. It measures how often the phase-randomized
reference records produce a Wolf exponent at least as large as that
obtained from the PIC record. If $\ensuremath{N_{\mathrm{PR}}}$ reference
records are generated and $N_{\mathrm{exceed}}$ of them satisfy
\begin{equation}
\lambda^{\mathrm{PR}}_{\max}\geq\lambda^{\mathrm{PIC}}_{\max},
\end{equation}
then the empirical probability $p$ is given by
\begin{equation}
p=\frac{1+N_{\mathrm{exceed}}}{1+N_{\mathrm{PR}}}.
\end{equation}
For the principal $M=0.5$ subsonic interval, one of the 32 phase-randomized
records exceeds the PIC value. Therefore, we have
\begin{equation}
p=\frac{1+1}{1+32}=\frac{2}{33}\simeq0.061,
\end{equation}
which means that the corrected empirical probability of obtaining
a PR reference exponent at least as large as the PIC value is approximately
$6.1\%$. Equivalently, the nominal one-sided confidence is 
\[
1-p\simeq0.939,
\]
or approximately $94\%$, just shy of the conventional confidence
level of $95\%$. Although this finite-time result does not establish
that the full PIC system occupies a stationary chaotic attractor,
it provides sufficient proof-of-principle evidence that the self-excited
IICSI dynamics with subsonic flow can support a chaotic basin. Because
the interval was selected from overlapping candidate windows, the
quoted confidence level should nevertheless be interpreted as indicative
rather than as a definitive global statistical measure.

\begin{figure}
\begin{centering}
\includegraphics[width=1\textwidth]{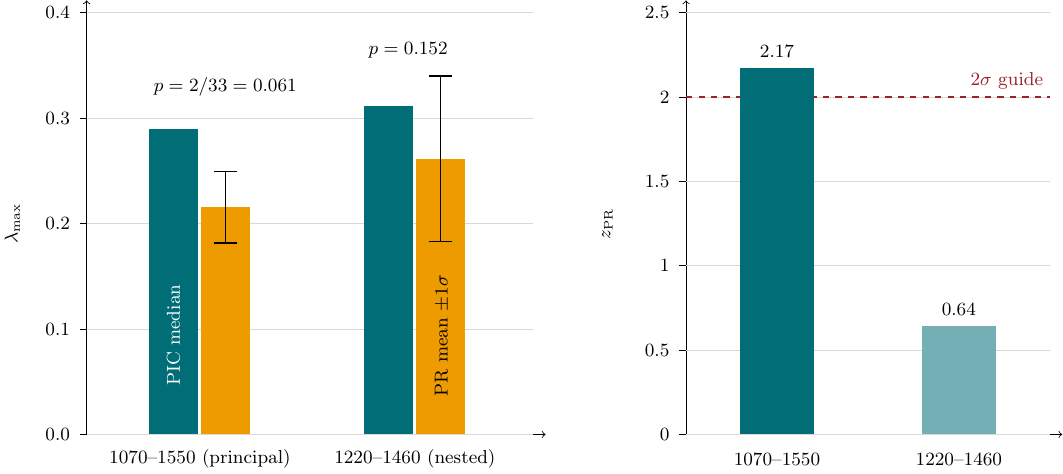}
\par\end{centering}
\caption{Summary of the Wolf-type chaos analysis for $M=0.5$. The PIC median
with the PR mean is shown in the left panel. The standardized PIC
excess is shown in the right panel. The best fit principal window
$W\omega_{pe}=[\ensuremath{1070,1550}]$ window reaches $z_{{\rm PR}}=2.17$
with empirical $p=2/33=0.061$. The nested window $W\omega_{pe}=[1220,1460]$
window has only $z_{{\rm PR}}=0.64$ and $p=0.152$, so it is positive
but not independently separated.}\label{fig:Summary-of-a}
\end{figure}

These results are also summarized in Fig.\ref{fig:Summary-of-a} for
the subsonic case of $M=0.5$, where in the left panel, the comparison
of the PIC median $\lambda_{\mathrm{max}}$ with the phase-randomized
(PR) mean $\pm1\sigma$ is shown and in the right panel, the standardizes
PIC excess of $\lambda_{\mathrm{max}}$ above the PR mean $z_{\mathrm{PR}}$
is shown
\begin{equation}
z_{{\rm PR}}=\left(\lambda^{{\rm PIC}}_{\max}-\overline{\lambda}^{{\rm PR}}_{\max}\right)/\sigma^{{\rm PR}}_{\lambda}.
\end{equation}

\section{Theory-to-PIC interpretation}

The theoretical and kinetic calculations address two complementary
parts of the proposed generic route to chaos. While the reduced model
establishes that an autonomous oscillator-charging-feedback system
possesses a bounded chaotic branch without any external periodic driving
mechanism. The PIC calculation then asks whether a self-consistent
kinetic plasma, including particle collection, photoemission, secondary-electron
emission (SEE) with surface regulation can locally reproduce the same
nonlinear feedback structure and exhibit an independently detectable
chaotic component.

For the subsonic case $M=0.5$, the interval 
\begin{equation}
1070\leq t\omega_{pe}\leq1550
\end{equation}
is selected by the weak-form regression as the interval giving the
strongest simultaneous compatibility with the theoretical charging
and oscillator equations. The sign-constrained fit satisfies the adopted
thresholds for both equations and reproduces the complete sign pattern
of the retained theoretical branch. Independently, the same interval
also gives the strongest separation between the PIC Wolf-type largest
Lyapunov exponent and the phase-randomized (PR) reference ensemble
with a nominal $94\%$ one-sided confidence level. Their concurrence
in the same finite-time regime is the proof-of-principle evidence
of the subsonic PIC simulation exhibiting a generic route to deterministic
chaos in self-excited plasma oscillations. However, we \emph{must
}be careful to note that the system does not show a global chaotic
attractor.

\subsection{Why the theoretical and PIC coefficients differ}

We note that the retained theoretical parameter vector is a numerically
verified member of a chaotic basin, not a quantitative prediction
derived from the particular PIC plasma parameters. Exact numerical
agreement between theoretical and PIC coefficients is therefore neither
expected nor required for the present proof of principle.

The dynamical origin of the mismatch is equally important and should
be noted. While the theoretical system is a three-variable autonomous
Markov closure with constant coefficients, the PIC system contains
kinetic degrees of freedom along with finite-particle noise. Eliminating
all these degrees of freedom would generally produce elaborate theoretical
model with state-dependent coefficients rather than an exact constant-coefficient
version of the reduced equations. A finite-window weak regression
would therefore estimate local effective coefficients and redistributes
unresolved degrees of freedom among correlated terms such as $\xi$,
$\dot{\xi}$, $z$, $\xi z$, and $z^{3}$. We would also not expect
that the PIC coefficients determined from the weak regression, when
inserted into the theoretical model, will produce a chaotic signature.
Direct substitution of the PIC-fitted coefficients into the theoretical
equations need not therefore produce the same attractor. Moreover,
integrating a coefficient vector obtained from one finite window indefinitely
extrapolates it beyond the regime in which it is inferred and can
fail to produce a global chaotic attractor in the higher-dimensional
PIC dynamics.

\section{Summary and Conclusions}

Starting from the Stuart--Landau description of a generic self-excited
ion-acoustic mode, we constructed a van der Pol-type modal coordinate
and coupled it to a finite-rate nonlinear charging variable. The resulting
autonomous system contains modal self-excitation, charge-dependent
frequency and force feedback, charging relaxation, phase-lagged modal
response, nonlinear mixed feedback, and saturation. Numerical bifurcation
analysis demonstrates a bounded chaotic basin can be achieved through
period doubling and band formation \cite{feigenbaum1978,ott2002}.
This establishes, at the reduced-model level, that self-excited plasma
oscillations can become chaotic without an externally imposed periodic
driver. The kinetic realization was tested with a one-dimensional,
open-boundary electrostatic PIC model of the debris-induced IICSI.
Particle collection, photoemission, electron-impact SEE, surface regulation,
and open-boundary reservoir dynamics provide a physically realizable
implementation of the charging-feedback loop. Weak-form regression
is used to test whether finite portions of the PIC trajectory project
onto the structure of the theoretical equations and the chaos diagnostics
were applied to the PIC observables themselves.

For the subsonic $M=0.5$ case, the interval $1070\leq t\omega_{pe}\leq1550$
passes the adopted fitting thresholds for both the charging and oscillator
equations and conforms to the complete sign pattern of the retained
theoretical branch. The same interval gives the strongest separation
of the PIC Wolf-type Lyapunov exponent from the phase-randomized reference
ensemble, with $p=0.061$ and a nominal confidence of about $94\%$.
Within the nested interval $1220\leq t\omega_{pe}\leq1460$, nonlinear
forecasting additionally shows that the debris potential and field
energy are more predictable than their phase-randomized references,
while the potential-well depth shows a weaker improvement and the
debris charge shows none. This forecasting result establishes nonlinear
deterministic organization in the collective plasma response, but
does not by itself establish chaos. The combination of structural
compatibility, nonlinear deterministic organization, and independent
trajectory-divergence evidence constitutes sufficient proof-of-principle
evidence that the self-excited IICSI can support a finite-time chaotic
regime. However, the supersonic $M=1.25$ case also displays nonlinear
deterministic organization, but it fails the charging-equation fitting
threshold and does not show a comparable Lyapunov-reference separation.
It is therefore more consistent with organized periodic or quasi-periodic
motion within the tested windows. This comparison is important because
it shows that the positive subsonic result is parameter-localized
and cannot be attributed automatically to PIC noise, spectral complexity,
or nonlinear predictability alone.

The reduced model proves the existence of an allowed autonomous route
and a bounded chaotic basin. The PIC calculation provides localized
kinetic support at a nominal confidence level of $94\%$. However,
it does not prove globally stationary PIC chaos, unique coefficient
identification, or exact equivalence between the full particle system
and the reduced model. Thus, not every self-excited plasma oscillation
is chaotic, but a physically realizable nonlinear charging-feedback
loop supplies a generic and accessible route by which such an oscillation
can become chaotic. The result of theoretical existence and the evidence
of finite-time kinetic dynamics together establish this route at the
proof-of-principle level.

\begin{appendices}

\section{Second order calculation}

Using the expression for $F_{1}$ from Eq.(\ref{eq:ch11-f1}), the
quadratic nonlinear term in Eq.(\ref{eq:ch11-f2}) can be written
as 
\begin{equation}
{\cal N}_{2}(F_{1},F_{1})=a^{2}{\cal N}_{2}(\varPhi,\varPhi)e^{2i\omega_{0}t}+|a|^{2}\left[{\cal N}_{2}(\varPhi,\varPhi^{*})+{\cal N}_{2}(\varPhi^{*},\varPhi)\right]+(a^{*})^{2}{\cal N}_{2}(\varPhi^{*},\varPhi^{*})e^{-2i\omega_{0}t}.
\end{equation}
For a symmetric ${\cal N}_{2}$, it reduces to 
\begin{equation}
{\cal N}_{2}(F_{1},F_{1})=a^{2}{\cal N}_{2}(\varPhi,\varPhi)e^{2i\omega_{0}t}+2|a|^{2}{\cal N}_{2}(\varPhi,\varPhi^{*})+(a^{*})^{2}{\cal N}_{2}(\varPhi^{*},\varPhi^{*})e^{-2i\omega_{0}t}.
\end{equation}
In general, Eq.(\ref{eq:ch11-f2}) can be written as 
\begin{equation}
(\partial_{t}-{\cal L}_{0})F_{2}=a^{2}G_{2}e^{2i\omega_{0}t}+|a|^{2}G_{0}+(a^{*})^{2}G_{-2}e^{-2i\omega_{0}t},
\end{equation}
with 
\begin{equation}
\begin{array}{lcl}
G_{2} & = & {\cal N}_{2}(\varPhi,\varPhi),\\
G_{0} & = & {\cal N}_{2}(\varPhi,\varPhi^{*})+{\cal N}_{2}(\varPhi^{*},\varPhi),\\
G_{-2} & = & {\cal N}_{2}(\varPhi^{*},\varPhi^{*}).
\end{array}
\end{equation}
As each of the harmonics is independent, we can solve the above equation
independently for each harmonic 
\begin{equation}
\begin{array}{lcl}
(\partial_{t}-{\cal L}_{0})F^{(2)}_{2} & = & a^{2}G_{2}e^{2i\omega_{0}t},\\
(\partial_{t}-{\cal L}_{0})F^{(0)}_{2} & = & |a|^{2}G_{0},\\
(\partial_{t}-{\cal L}_{0})F^{(-2)}_{2} & = & (a^{*})^{2}G_{-2}e^{-2i\omega_{0}t}.
\end{array}
\end{equation}
For each harmonic, we seek a solution for $F_{2}$ 
\begin{equation}
\begin{array}{lcl}
F^{(2)}_{2} & = & a^{2}\varPhi_{2}e^{2i\omega_{0}t},\\
F^{(0)}_{2} & = & |a|^{2}\varPhi_{0},\\
F^{(-2)}_{2} & = & (a^{*})^{2}\varPhi_{-2}e^{-2i\omega_{0}t},
\end{array}
\end{equation}
combining which, we get a general form of solution for $F_{2}$ as
\begin{equation}
F_{2}=a^{2}\varPhi_{2}e^{2i\omega_{0}t}+|a|^{2}\varPhi_{0}+(a^{*})^{2}\varPhi_{-2}e^{-2i\omega_{0}t},
\end{equation}
where $\varPhi_{-2}=\varPhi^{*}_{2}$. The amplitude $a$ is essentially
a constant at this order, which is why we don't see any term $\propto\partial_{t}a$.

\section{Resonant harmonic}

We note that the resonant eigenfunction is 
\begin{equation}
F\sim\varPhi e^{\pm i\omega_{0}t}.
\end{equation}
To prove it, we assume that there exists such an eigenfunction at
order ${\cal O}(\epsilon^{2})$, so that 
\begin{equation}
(\partial_{t}-{\cal L}_{0})F_{2}=c\varPhi e^{i\omega_{0}t}.
\end{equation}
This equation requires solution for $F_{2}$ of the form 
\begin{equation}
F_{2}\sim ct\,\varPhi e^{i\omega_{0}t},
\end{equation}
which is a secular term and perturbations grow indefinitely. Note
that $e^{-i\omega_{0}t}$ is just the complex conjugate of the positive
term and necessarily both harmonics would be present in the solution
as the coefficients of the equation are real. 

\section{Expression for $G_{21}$}

We note that $G_{21}$ is the coefficient of the terms proportional
to $a|a|^{2}e^{i\omega_{0}t}$ in the expression
\begin{equation}
2{\cal N}_{2}(F_{1},F_{2})+{\cal N}_{3}(F_{1},F_{1},F_{1}).
\end{equation}
Writing $\varepsilon=e^{i\omega_{0}t}$, we have the expressions for
$F_{1,2}$ as
\begin{eqnarray}
F_{1} & = & a\varPhi\varepsilon+a^{*}\varPhi^{*}\varepsilon^{-1},\\
F_{2} & = & a^{2}\varPhi_{2}\varepsilon^{2}+|a|^{2}\varPhi_{0}+(a^{*})^{2}\varPhi_{-2}\varepsilon^{-2}.
\end{eqnarray}
The expression for ${\cal N}_{2}$ is then
\begin{eqnarray}
{\cal N}_{2}(F_{1},F_{2}) & = & a^{3}\varepsilon^{3}{\cal N}_{2}(\varPhi,\varPhi_{2})+a|a|^{2}\varepsilon{\cal N}_{2}(\varPhi,\varPhi_{0})+a(a^{*})^{2}\varepsilon^{-1}{\cal N}_{2}(\varPhi,\varPhi^{*}_{2})\\
 &  & +\,a^{*}a^{2}\varepsilon{\cal N}_{2}(\varPhi^{*},\varPhi_{2})+a^{*}|a|^{2}\varepsilon^{-1}{\cal N}_{2}(\varPhi^{*},\varPhi_{0})\nonumber \\
 &  & +\,(a^{*})^{3}\varepsilon^{-3}{\cal N}_{2}(\varPhi^{*},\varPhi^{*}_{2}).\nonumber 
\end{eqnarray}
As only the second and the fourth terms contain $\varepsilon=e^{i\omega_{0}t}$,
we have the contributions to $G_{21}$ from $2{\cal N}_{2}$ as
\begin{equation}
G^{{\cal N}_{2}}_{21}=2\left[{\cal N}_{2}(\varPhi,\varPhi_{0})+{\cal N}_{2}(\varPhi^{*},\varPhi_{2})\right]\label{eq:g21n2}
\end{equation}
as $a^{*}a^{2}=a|a|^{2}$.

Similarly, assuming that the trilinear operator ${\cal N}_{3}$ is
symmetric in its arguments, for ${\cal N}_{3}(F_{1},F_{1},F_{1})$,
we have
\begin{equation}
{\cal N}_{3}(U+V,U+V,U+V)={\cal N}_{3}(U,U,U)+3{\cal N}_{3}(U,U,V)+3{\cal N}_{3}(U,V,V)+{\cal N}_{3}(V,V,V),
\end{equation}
where we have written $F_{1}=U+V$, with $U=a\varPhi\varepsilon$
and $V=a^{*}\varPhi^{*}\varepsilon^{-1}$. Among all these terms,
the positive fundamental harmonic $e^{i\omega_{0}t}$ comes only from
the term ${\cal N}_{3}(U,U,V)$
\begin{equation}
{\cal N}_{3}(U,U,V)=a|a|^{2}\varepsilon{\cal N}_{3}(\varPhi,\varPhi,\varPhi^{*}).
\end{equation}
and thus the contributions to $G_{21}$ from ${\cal N}_{3}$ is
\begin{equation}
G^{{\cal N}_{3}}_{21}=3{\cal N}_{3}(\varPhi,\varPhi,\varPhi^{*}).\label{eq:g21n3}
\end{equation}
Combining Eqs.(\ref{eq:g21n2},\ref{eq:g21n3}), we have
\begin{equation}
G_{21}=2{\cal N}_{2}(\varPhi,\varPhi_{0})+2{\cal N}_{2}(\varPhi^{*},\varPhi_{2})+3{\cal N}_{3}(\varPhi,\varPhi,\varPhi^{*}).
\end{equation}

\section{Projection coefficients}

We start with residual of the expansion given by Eq.(\ref{eq:expansion})
\begin{equation}
\delta X_{\perp}(t)\equiv r(t;A_{X},B_{X})=\delta X(t)-A_{X}\xi(t)-B_{X}\dot{\xi}(t).
\end{equation}
We now minimize the squared residual
\begin{equation}
J(A_{X},B_{X})=\int^{t_{2}}_{t_{1}}r^{2}(t;A_{X},B_{X})\,dt
\end{equation}
within the finite time window $[t_{1},t_{2}]$. At the least-square
minimum, we must have
\begin{equation}
\frac{\partial J}{\partial A_{X}}=\frac{\partial J}{\partial B_{X}}=0.
\end{equation}
This immediately shows that
\begin{equation}
\left\langle \xi,r\right\rangle =\left\langle \dot{\xi},r\right\rangle =0,
\end{equation}
where
\begin{equation}
\left\langle a,b\right\rangle =\int^{t_{2}}_{t_{1}}a(t)b(t)\,dt
\end{equation}
is the inner product within that finite time interval. As $r$ is
orthogonal to $\xi$ and $\dot{\xi}$, we have used the the symbol
$\delta X_{\perp}$ to indicate that the remainder is orthogonal.

Applying the orthogonality condition, we have
\begin{eqnarray}
\left\langle \xi,\delta X-A_{X}\xi-B_{X}\dot{\xi}\right\rangle  & = & 0,\\
\left\langle \dot{\xi},\delta X-A_{X}\xi-B_{X}\dot{\xi}\right\rangle  & = & 0,
\end{eqnarray}
which simplifies to
\begin{eqnarray}
A_{X}\langle\xi,\xi\rangle+B_{X}\langle\xi,\dot{\xi}\rangle & = & \langle\xi,\delta X\rangle,\\
A_{X}\langle\dot{\xi},\xi\rangle+B_{X}\langle\dot{\xi},\dot{\xi}\rangle & = & \langle\dot{\xi},\delta X\rangle.
\end{eqnarray}
As $\langle\dot{\xi},\xi\rangle=\langle\xi,\dot{\xi}\rangle$, we
can solve the above equations to obtain
\begin{equation}
\begin{pmatrix}\langle\xi,\xi\rangle & \langle\xi,\dot{\xi}\rangle\\
\langle\xi,\dot{\xi}\rangle & \langle\dot{\xi},\dot{\xi}\rangle
\end{pmatrix}\begin{pmatrix}A_{X}\\
B_{X}
\end{pmatrix}=\begin{pmatrix}\langle\xi,\delta X\rangle\\
\langle\dot{\xi},\delta X\rangle
\end{pmatrix},
\end{equation}
from which we can obtain the expressions for $(A_{X},B_{X})$ as
\begin{eqnarray}
A_{X} & = & \Delta^{-1}\left(\langle\xi,\delta X\rangle\langle\dot{\xi},\dot{\xi}\rangle-\langle\dot{\xi},\delta X\rangle\langle\xi,\dot{\xi}\rangle\right),\\
B_{X} & = & \Delta^{-1}\left(\langle\dot{\xi},\delta X\rangle\langle\xi,\xi\rangle-\langle\xi,\delta X\rangle\langle\xi,\dot{\xi}\rangle\right),
\end{eqnarray}
where
\begin{equation}
\Delta=\langle\xi,\xi\rangle\langle\dot{\xi},\dot{\xi}\rangle-\langle\xi,\dot{\xi}\rangle^{2}.
\end{equation}
For this analysis to be valid, we need $\Delta\neq0$. If however
$\Delta=0$, both $(\xi,\dot{\xi})$ are linearly dependent over that
particular time interval and their separate contributions cannot be
uniquely determined.

\bmhead{Acknowledgements}

This work is carried out under the research grant RAC-S/GU/2024/4/74 of ISRO Regional Academic Centre for Space, Gauhati University.

\section*{Declarations}

\begin{itemize}
\item {\bf Funding} This work is supported by ISRO's RESPOND programme under the research grant RAC-S/GU/2024/4/74 of ISRO Regional Academic Centre for Space, Gauhati University.

\item {\bf Competing interests} The authors declare no competing interests.
\item {\bf Ethics approval} Not applicable
\item {\bf Consent for publication} Not applicable
\item {\bf Data availability} Not applicable
\item {\bf Materials availability} Not applicable
\item {\bf Code availability} https://doi.org/10.5281/zenodo.24225095
\item {\bf Author contribution} MPB conceptualised the idea, did the initial research, worked out detailed methodology, and written the first draft. BJ  did all simulations. AG took part in initial formative conceptualisation or the work. KB supervised the work. The final version was approved all the authors.

\end{itemize}

\end{appendices}

\bibliographystyle{sn-mathphys-num}
\bibliography{chaos-rmpp}

\end{document}